# CURRENT TRENDS AND FUTURE RESEARCH DIRECTIONS FOR INTERACTIVE MUSIC

MAURICIO TORO

Universidad Eafit, Department of Informatics and Systems, Colombia

E-mail: mtorobe@eafit.edu.co

**Abstract**

In this review, it is explained and compared different software and formalisms used in music interaction: sequencers, computer-assisted improvisation, meta-instruments, score-following, asynchronous dataflow languages, synchronous dataflow languages, process calculi, temporal constraints and interactive scores. Formal approaches have the advantage of providing rigorous semantics of the behavior of the model and proving correctness during execution. The main disadvantage of formal approaches is lack of commercial tools.

## 2 Introduction

Technology has shaped the way on which we compose and produce music: Notably, the invention of microphones, magnetic tapes, amplifiers and computers pushed the development of new music styles in the 20th century. In fact, several artistic domains have been benefiting from such technology developments; for instance, *Experimental music*, *non-linear music*, *Electroacoustic music*, and *interactive music*.

*Experimental music* is composed in such a way that its outcome is often unforeseeable; for instance, it may contain random generated tones, computer-generated content, variable-duration notes and "free" content. It may also include atonal melodies and microtones.

Another domain is *non-linear music*, in which the scenario is divided in parts whose order can be chosen at execution time. We will use the term "non-linear" music in that sense. Non-linear music exists from many centuries ago; for instance, Mozart's minuets in which the order of work's musical material was determined by coin-tosses.

*Electroacoustic music* was originated by the incorporation of electronic sound production into compositional practice. It subsumes styles such as *musique concrète* (French for *concrete music*), *Acousmatic music*, *musique mixte* (French for "mixed" music) and *Electronic music*. Note that Electroacoustic and Experimental music are not mutually exclusive: a piece can belong to both styles or to a single one, for instance, Experimental music explores composition with microtones which does not incorporate electronic sounds.

*Interactive music* deals with the design of scenarios where music content and interactive events are handled by computer programs. Examples of such scenarios are music art installations, interactive museum exhibitions, some Electroacoustic music pieces, and some Experimental music pieces. In Table 2, it is presented a literature mapping of the different mathematical models and software that will be presented in this article.

| | |
|---|---|
| Sequencers | Pro Tools, Qlab, Ableton Live |
| Computer-assisted improvisation | [12, 48, 61] |
| Meta-instruments | [39] |
| Score following | [23] |
| Asynchronous dataflow languages | [88] |
| Synchronous dataflow languages | [37, 38, 33, 17, 41] |
| Process calculi | [60, 59, 80, 78, 79, 75, 5, 100, 58, 59, 105] |
| Temporal constraints | [1, 51, 20] |
| Interactive scores | [4, 105, 104, 59, 106] |

Table 2: Literature mapping of mathematical models and software for music interaction

In what follows we briefly explain Experimental music, non-linear music, Electroacoustic music and interactive music. In this thesis we will focus on interactive music. We are interested in Electroacoustic, Experimental and non-linear music that is interactive. In this section, we introduce the problems that arise when designers and composers want to write a score for interactive music, and the problems with existing computer tools to compose and perform interactive music; afterwards, we briefly describe some background concepts and we propose a solution based on the formalism of interactive scores.

In this section we briefly define Experimental music, non-linear music, Electroacoustic music and interactive music. To clarify the classification of these domains, we present a Venn's diagram in Figure 1: The diagram shows the intersection between the different domains. Figure 1 includes *music art installations*, which are an interesting subset of interactive music; and *Tape music*, which is a subset of Electroacoustic music that is linear (i.e., parts have a fixed order) and is not interactive.

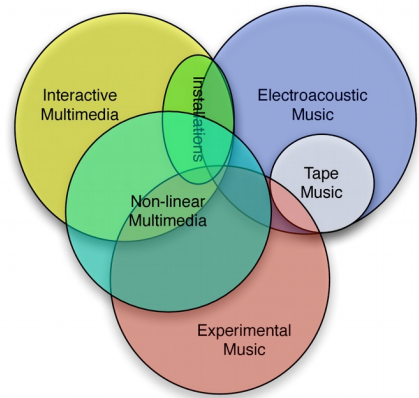

Figure 1: Intersection between Electroacoustic music, non-linear music, Experimental music and interactive music.

## Experimental music.

Nyman argues that, in Experimental music, a score may no longer represent a sound by the means of western music notation [57]: Composers may provide the performer the means of making calculations to determine the nature, timing and spacing of sounds. Composers may indicate temporal areas in which a number of sounds may be placed. Experimental music can span from a minimum of organization to a minimum of arbitrariness. As an example, Christopher Hobb's *voicepiece (1967)* is written for any number of vocalists and any length. Nyman argues that, usually, in Experimental music pieces, certain time frames may be chosen at random and filled with sounds.

Nyman argues that an important feature of Experimental music is the diversity of processes available; processes may be relationships between chance and choice. He argues that there are five types of processes: (1) *change determination processes*; for instance, when Cage used random numbers to choose tones, and also when he wrote pieces in which it was required to take information from the telephone directory during performance; (2) *people processes*, for instance, the eventuality of players getting lost or an unknown number of players; (3) *contextual processes*, such as actions taken on unpredictable conditions within the musicians or the audience; (4) *repetition processes*, such as unbounded loops; and (5) *electronic processes*, difficult to describe because they are not well formalized.

A characteristic of Experimental music is that, often, the starting and ending times of a piece are unknown. As an example, Nyman argues that in Wolff's *duo II for pianists (1958)*, the beginning and the ending times are determined in performance by the circumstances of the concert occasion. As another example, Nyman discussed Reich's *pendulum music (1968)*. In this piece, microphones are suspended from the ceiling. The piece begins when the performers swing the microphones and turn on the amplifiers; the piece ends after all microphones come to rest.

Nyman argues that performing Experimental music goes above and beyond performing of *Western music* because of all the possibilities that can be modeled with the five types of processes, and the unknown starting and ending times of a piece, as explained above.

**Non-linear music.**

Since 1950, computer technology is used to control sound structures; however, there is a long history of non-linear music in western culture. Vickery argues that, in the 20th century, there are examples of non-linear music such as Boulez's *third piano sonata (1958)*, and free improvisation with game strategies such as interactive electronics from Gordon Mum and several Stockhausen's pieces. Nonetheless, such an interest is not new. In fact, Vickery argues that Mozart composed minuets and trios in which the order of work's musical material was determined by coin-tosses, as we stated before.

Vickery has composed some non-linear pieces [115] in the 21st century. As an example, *ladders and snakes (2000)* is a piece in which the *ladder* processes descend to improvise in a later section, and the *snake* processes ascend to an earlier section, as a flash back in a film. As another example, *splice (2002)* is a piece in which the computer performs meta-music shaping of the sound made by the musician. Finally, in Vickery's piece *parallel trajectories (2003)*, performers have a score map with different paths from start to end, and they can also choose to stay silent in some parts. As an example, the score is presented in Figure 2.

0.2

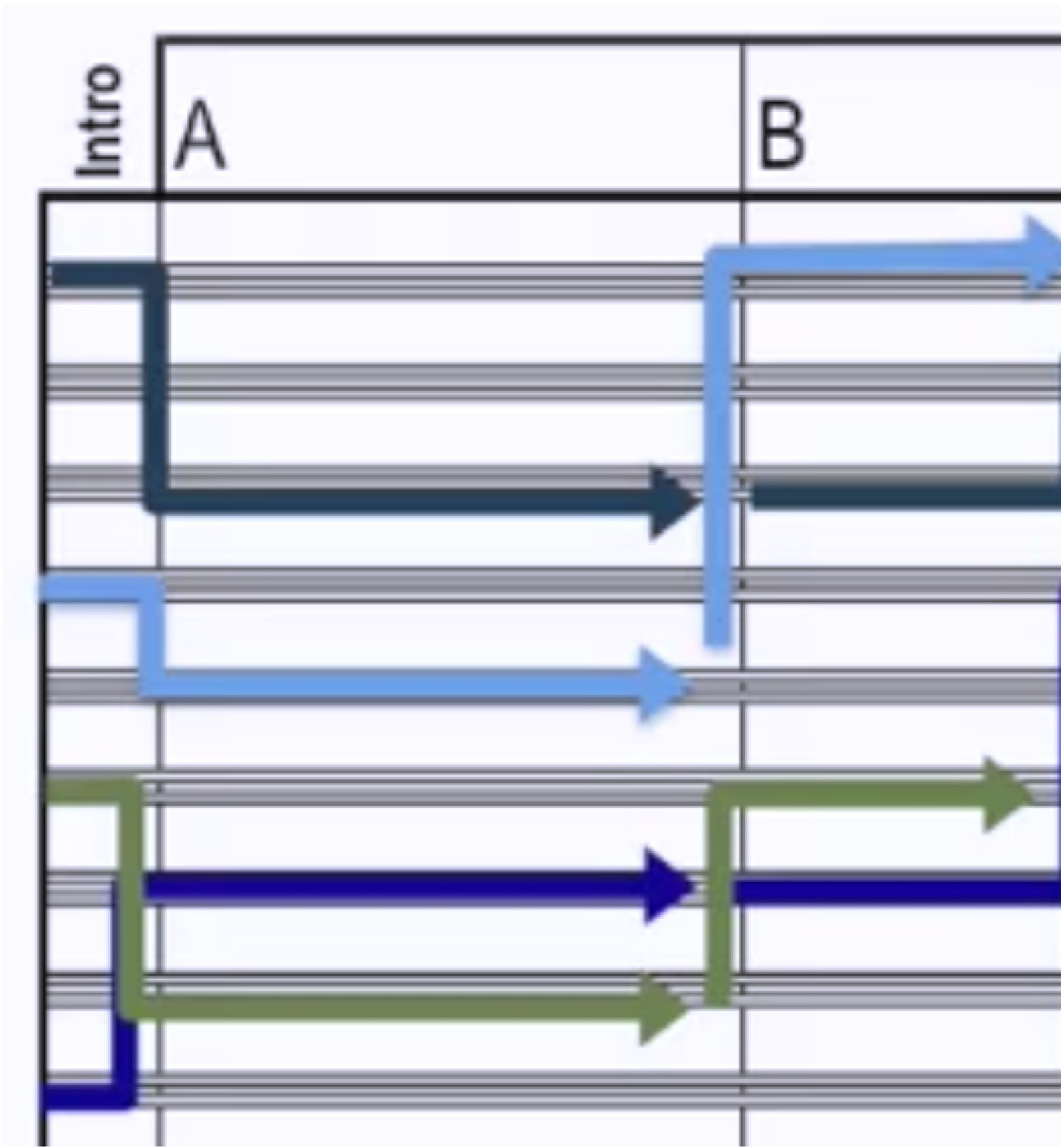

Figure 2: Score of Vickery's *parallel trajectories (2003)*. There are 14 lines of musical material and each of the 9 players is provided with four of the lines. There are 9 "modal points" in the score in which the player may choose a different line or choose to remain silent until the next point.

Furthermore, Vickery argues that computer coordination of live musical performance allows for the control and synchronization of the score; for instance, non-linear selection of music material [117]. Music is traditionally linear: left-to-right and top-to-bottom. Computer music offers two main new possibilities according to Vickery: (1) Permutation of large structural blocks of music such as Stockhausen's *momente (1962)*, and (2) interactive generative processes may be used in real-time. There are some other implications of such a computer-controlled behavior, according to Vickery [117]. As an example, Jason Freeman's *glimmer (2004)* is written for chamber orchestra and audience participation by waving four-inch LED sticks. Vickery's *delicious ironies (2002)* has also an unpredictable environment for the solo improviser with sample choice, playback speed, duration, volume and pan. As another examples, Vickery recalls Stockhausen's spectral analysis used in *zyklus (1959)* and *regrain (1959)*.

According to Vickery, non-linearity allows pieces to have *openness of interpretation* and *openness of content* [116]. Vickery cites some interesting examples. Game based analysis first used by Xenakis in *duel (1959)* and *strategies (1962)*, then used by John Zorn in *cobra (1984)*, allows the musician to give commands to games. Richard Teitelbaum, creator of *automata (1978)*, presents an analogy to finite state automata in which a system responds to user actions. The californian group *The HUB* is a computer network band in which the musicians and sounds communicate through a network.

Although the many examples that Vickery explained in his articles, he argued towards the urgent need of symbiotic human-machine interactive software to compose non-linear music [116]. In fact, we argue in this section why Vickery's preoccupation can be extended to non-linear music in general, for instance, in music art installations.

## Electroacoustic music.

All Electroacoustic music is made with electronic technology. Some electroacoustic compositions make use of sounds not available in typical acoustic instruments, such as those used in a traditional orchestra. Some Electroacoustic music can be created using non-acoustic technology that exists only in a recorded format (as a fixed medium), and is composed for reception via loudspeakers. The compositional material is not restricted to the inclusion of sonorities derived from musical instruments or voices, nor to elements traditionally thought of as "musical" (e.g., melody, harmony, rhythm and meter), but rather admits any sound, acoustic or synthetic. With the aid of various technologies, such as tape recorders and digital signal processing tools, this material can then be combined, juxtaposed, and transformed, in any conceivable manner [1].

A form of Electroacoustic music, specifically composed for loudspeaker presentation, is *Acousmatic music*. Unlike scored music, compositions that are purely acousmatic exist solely as audio recordings. The term *acousmatic* was introduced by Pierre Schaeffer and refers to the listening experience of *concrete music* in which the audience hears the music from the loudspeakers, without seeing the source of the sound [2]. In an acousmatic concert, the sound component is produced using pre-recorder media, or generated in real-time using a computer. The work is often *diffused* by the composer (if present), but the role of the interpreter can also be assumed by another musician. The main role of musician is to control *spatialisation*. As an example, consider one of Schaeffer's earliest work *five studies of noises (1948)* made without a computer.

1 http://en.wikipedia.org/wiki/Electroacoustic\_music
2 en.wikipedia.org/wiki/Acousmatic_music .

The term *concrete music* is defined by Schaeffer as an opposition with way musical work usually goes. Instead of notating musical ideas on a paper with the symbols of solfège and entrusting their realization to well-known instruments, the question is to collect concrete sounds, wherever they came from, and to abstract the music values they were potentially containing. According to Pierre Henry, another well-known composer of this style, concrete music was not a study of timbre, it is focused on envelopes and forms[3].

A subtype of concrete music, in which sound was registered in magnetic tapes, is called *Tape music*[4]. In such a style, the starting and ending times of all the sounds remain fixed once the composition is over; as opposed, to some pieces of *acousmatic music* in which there is real-time sound generated by computer which order may change.

There is another style subsumed by Electroacoustic music: *"Mixed" music*, which merges acoustic sounds from traditional instruments played by musicians with electroacoustic sounds (diffused by loudspeakers). As an example, in Manoury's *partita I (2006)* for solo viola and live electronic effects, in Section VIIC, the composer wrote a note indicating that the all parts have to be played but in any order. The order is chosen by the musician. This is an example of non-linearity in Electroacoustic music. Another well-known example of "mixed" music is Manoury's *pluton (1988)* for piano and live electronics, and Stockhausen's *mikrophonie I (1964)* for tam-tam, microphone and filters.

## Interactive music.

Interactive music deals with the design of scenarios where music content and interactive events can be handled by computer programs. Designers usually create music content for their scenarios, and then bind them to external interactive events controlled by *Max/MSP* or *Pure Data (Pd)* programs [70, 68]. We recall that examples of interactive music are interactive museum exhibitions and music installations.

*music art installations* are an artistic genre of three-dimensional works that are often site-specific and designed to transform the perception of a space. Installations evolved through the use of new and ever-changing technologies: from simple video installations, they expanded to include complex interactive, music and virtual reality environments. Interactive installations were most frequently created and exhibited after 1990s, when artists were particularly interested in using the participation of the audiences to co-author the meaning of the installation[5]. As an example, there is an interactive installation based on spatial sensing written in Max [119]. Another example is an interactive installation based on probabilistic control [14]. Both installations are non-linear in the sense that the order in which they diffuse video and sound is unforeseen and depends on user interactions.

In addition to Max, interactive music scenarios are also designed with commercial sequencers. Commercial sequencers for interactive music are based on a fixed timeline with a very precise script such as *Pro Tools*[6], or a more flexible script using cue lists, for instance, the theater cue manager *Qlab*[7]. Another software to design such scenarios is *Ableton Live*[8]. Live is often used in Electroacoustic music and performing arts.

**Example 1** *Figure 3 shows the user interfaces of cue lists and timeline based sequencers, respectively.*

*0.45*

## Cue Lists

| Cue No | Next Cue | Description | Status | Trigger | Channels | Effects |
|--------|----------|-------------|--------|---------|----------|---------|
| 0 | 1 | Starting Cue | 0% | Manual | | None |
| 1 | 2 | Cue Number 1 | 0% | Manual | 1,2,3,4,5,2... | blue and |
| 2 | 3 | Cue Number 2 | 0% | Manual | 6,7,8,9,10 | None |
| 3 | 4 | Cue Number 3 | 0% | Manual | | None |
| 4 | 5 | Cue Number 4 | 0% | Manual | 1,2,3,4,5,6... | None |
| 5 | 6 | Cue Number 5 | 0% | Manual | 37,60,70,2... | None |
| 6 | 0 | Cue Number 6 | 0% | Manual | 1,11,12 | None |

[ ] Find next    Find previous    Highlight    ☐ Match case

| Options ▾ | Cue Options ▾ | Run Cue | Add Cue | Delete Cue | Selected Cue |

| | | | | | | |
|---|---|---|---|---|---|---|
| ◉ ▾  →► a...z ↓↑•‖•  ≡•‖•  □→⅋  Grid ♪ | 0\|0\|240 ▾ | Nudge ♫ | 0\|0\|060 ▾ | Cur |
| Bars:Beats | 1 | 17 | 33 | 49 | 65 |
| Min:Secs | 0:00 | 0:30 | 1:00 | 1:30 | 2:00 | 2 |
| Time code | 01:00:00:00 | 01:00:30:00 | 01:01:00:00 | 01:01:30:00 | 01:02:00:00 | 0 |
| Feet+Frames | 0+00 | | 90+00 | | 180+00 | |

*Figure 3: Cue-list based Qlab (above) exhibits a list on events and associated actions; it also defines whether and event is triggered by the computer or by the user. Timeline sequencer Protools (below) exhibits a timeline with several sound objects; starting and ending times are fixed and cannot be changed during performance.*

Another well-known fixed timeline sequencer is the *Acousmograph* which is a software to represent graphically sounds in a composition. In fact, the acousmograph has been used by Pierre Couprie for musicological analysis [25]. It is also worth to note that the acousmograph has been used to represent Gyorgy Ligeti's *artikulation (1958)*, as shown in Figure 4[9].

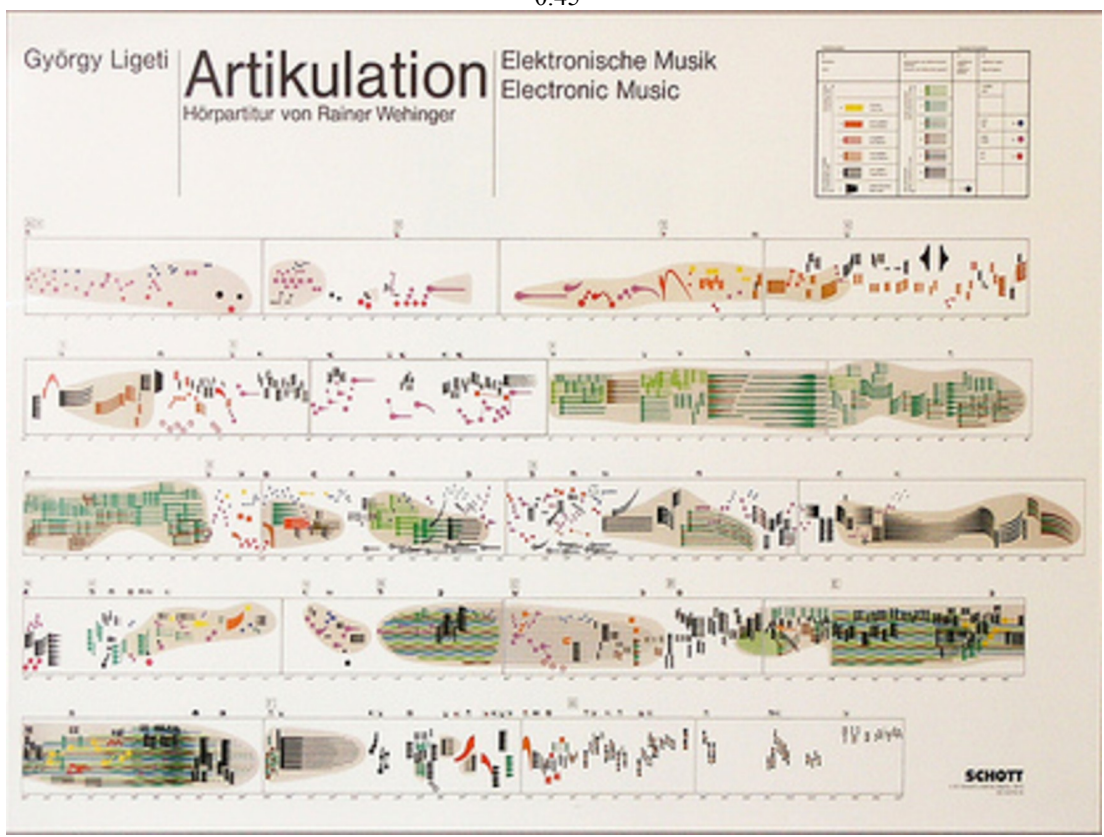

Figure 4: Visual listening score of Gyorgy Ligeti's *artikulation (1958)* created by Rainer Wehinger using *acousmograph*.

9 A video can be found at http://wn.com/artikulation_ligeti.

In what follows, we define the fixed timeline and the cue-lists time models, and the problems that have arisen because of the duality between these two time models, among other problems.

# 3 Current problems with Interactive music Scenarios

We have identified seven problems with existing software to design multimedia scenarios: (1) there is no formal model for multimedia interaction, (2) multimedia scenarios have limited reusability and difficulties with the persistence of multimedia scenarios, (3) time models (fixed timeline and cue lists) are temporally unrelated, (4) most multimedia interaction software products provide no hierarchy, (5) the different time scales are unrelated, (6) schedulers for multimedia scenarios are not appropriate for soft real-time, and (7) there is no model to combine temporal relations and conditional branching.

The main problem with interactive music scenarios is that there are two different time models, but existing tools only use one, and tools that allow both, offer both time models temporally unrelated. To understand this problem, we must travel 2500 years back in time. Desainte-Catherine *et al.* argued that this problem was already discussed by Parmenides of Elea and Heraclitus of Ephesus long before the invention of computers [28] .

## Problems with the time models.

According to Desainte-Catherine *et al.*, what we call today Tape music, that began by editing and mixing sounds in magnetic tapes, is composed in a writing-oriented manner that corresponds to the *arrow metaphor* discussed by Parmenides. Parmenides argued that there are eternal properties and ordered events; for instance, "Socrates was born before he died". According to Parmenides, timeline goes from past to future. In this paradigm, it is difficult to define changes in the objects in the timeline. In fact, the only changes allowed at performance time of Tape music are in pan, volume, spacialization, among others parameters, but not on the starting and ending time of the sounds nor individual parameters for each sound. In contrast, many pieces of Experimental and Electroacoustic music, are based on real-time sound synthesis. They are usually written in *asynchronous dataflow languages* such as Max. According to Desainte-Catherine *et al.*, interactive music is performance-oriented, and, for that reason, music objects and time representation are quite poor. Performance-oriented software corresponds to the *river metaphor* described by Heraclitus: "we never bath twice in the same river". In this paradigm, the inference of the events flows is from the future, backwards because events are being "scheduled".

Identity is hard to define in Heraclitus' paradigm; for that reason, according to Desainte-Catherine *et al.*, we cannot define a permanent environment in asynchronous dataflow languages such as Max/MSP [70]. Time-stamped data is handled as a queue and there is only available a limited timeline to schedule the triggering of static events in most asynchronous dataflow languages. Nonetheless, it is worth noticing the effort made my Puckette to include a timeline in Pure Data, as shown in Figure 5 [69].

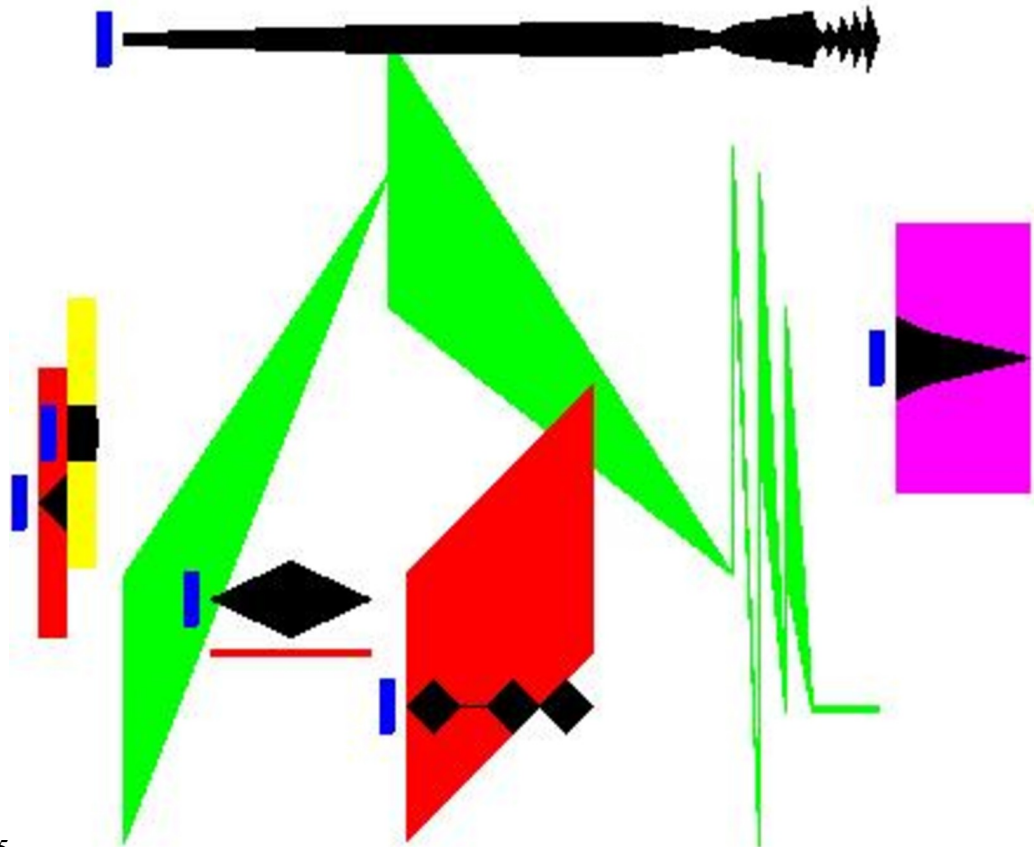

0.65

Figure 5: Writing a score in Pure Data. Horizontal axis represents time and the vertical axis frequency. Objects represent Pure Data's data structures. Shapes and colors are chosen by the composer to represent the data structures of the piece.

The problem of identify is important for both Electroacoustic and Experimental music. One implication is the ownership of Electroacoustic music, as explained by Dahan *et al.* [26]. According to Nyman, Cardew argued that when we hear on a tape or disk is indeed the same playing, but divorced from its natural context. As an example, Nyman argued that David Tudor (pianist) played Cage's *4'33" (1952)* and people think that Cage's *4'33" (1952)* is a piece for piano, but it is a piece that can be played by the means of any instrument [57].

**Problems with time scales.**

In addition to the problem of identity, Schwer discussed another philosophical problem related to linguistics [86], which we believe that it is also related to music: Aristotle argued that between two time instants there is always a time instant. Therefore, the metaphoric timeline seams like the set of real numbers. Nonetheless, according to Schwer, there is a discrete understanding of time in Physics; for instance, in *quantic mechanics*, *Planck's time* is the smallest measure of time  seconds; in the atomic clock is  seconds; however, humans only discriminate at  seconds.

In Computer Science, as in Physics, time is also discrete because it is defined by the occurrence of events. For events to occur they have to be observed and this is discrete in nature. In favor of discrete time, the Stoics argued that the set of atomic instants is a discrete structure, thus we can pass from one instant to the next instant, according to Schwer.

The duality between discrete and continuous time is also a problem in music interaction when we think about all the time scales available; for instance, user gestures, control events, sound processing and video processing. All those processes work at different time scales, and they are usually unrelated one from another in existing tools. music signals are continuous when they are analogic. Once they are sampled into the computer, they become discrete; however, they can be though of continuous in the sense that a listener will perceive them as continuous. In contrast, control signals, used to synchronize different media, are discrete time, and they are also perceived as discrete by the listeners.

## Problem with synchronization.

There is another problem derived from the time scales, as we discussed in [103]. The description of a music scenario requires a consistent relationship between the representation of the scenario in the composition environment and the execution. Artistic creation requires a composition of events at different time scales. As an example, it is easy to describe that a video begins when the second string of a guitar arpeggio starts, but how can we achieve it in practice if the beginning of the notes of the arpeggio is controlled by the user?

The problem emerges at runtime. The example given above is very simple, but under high CPU load, a system interruption at the point of playing the arpeggio and the video can often lead to desynchronization, which is the case with Pure Data and Max. Usually, these eventualities are not considered by developers, as the quality of systems is evaluated according to an average performance. Nonetheless, during performance, it is desired that the system works well even under high CPU load, which is common when these systems process sound, video and image simultaneously.

The synchronization between the arpeggio and the video must be achieved in every execution. If it does not work for a performance, concert or show, the system performance is not satisfactory. Usually, artists prefer that an event is canceled if the event is not going to be properly synchronized with all the other media. Most users want a system that ensures that the events are either launched as they were defined in the score or they are not produced. Another alternative is based on the synchronization strategies for score following systems proposed by Echeveste *et al.* [30]. Echeveste's strategies are designed to define behaviors for the cases in which events are not always properly synchronized with other media due to musician's mistakes during performance or due to incorrect tempo calculations by the score following system.

Interactive music belongs to the realm of *soft real-time*. We argue that in soft real-time, the usefulness of a result degrades after its deadline, thereby degrading the system's quality of service; whereas in *hard real-time* missing a deadline is a total system failure (e.g., flight control systems). It is difficult to ensure determinism in the execution of music processes (e.g., sound, video and 3D images) in the soft real-time realm. Some *hard real-time* operating system like *RT Linux*[10] or *RedHawk*[11] include priority queues for processes to respect hard real-time constraints; however, in common operating systems, the user does not have this type of control. Note that software like Max and Live do not work on Linux.

10 http://www.windriver.com/index.html
11 http://real-time.ccur.com/concurrent_redhawk_linux.aspx

## Problems with conditional branching.

Another issue arises when we think of non-linear music. When we think about choices based on conditions, we must consider *causality*. Causal relation is studied by *metaphysics*. According to Keil, substances are not causes; for instance, "if knife then always wound" is incorrect: An event and a verb are missing [42]. In interactive music, "If note 1 then always note 2" is also incorrect. A causal relation could be "when note 1 starts, then note 2 starts","whenever note 1 ends the note 2 ends", or "when the note 1 gets to a volume peak, then note 2 starts"; however, most tools do not provide this kind of causal relations. Keil explains that physical systems are described in non-perturbed situations, but such rules may not always apply in real-life situations. As an example, a fire match will not light without oxygen, although a cause of lighting a match is to rub it against a striker. For that reason, when we model non-linear music, we must consider user interactions. We must also consider that these interaction may arrive at any time.

Keil also points out that an event always has different causes susceptible of exceptions because the causes include less than the total state of the universe. For that reason, the causal relation is not transitive; therefore, the flapping of a butterfly's wing is not the cause of a storm on the other side of the world, according to Keil. As a consequence, we argue that users' choices should be made over single temporal objects (e.g., sounds or videos), instead of sequences of temporal objects. To choose a sequence of temporal objects, the sequence should be contained in one temporal object. In conclusion, each object must know who was its direct cause. In both figures, there is a mutually exclusive choice between two objects. If a composer wants to write a choice between two sequences of two objects, each two-object sequence must be contained inside a bigger object.

Up to now we have considered causality dissociated from time, as treated by Keil; however, Russel gives a definition of causality that includes a time interval: "Given an event $e$, there is an event and a time interval $\tau$, such that, every time that occurs, it is followed by , after such an interval has passed" [81]. We believe that Russels' definition is appropriate for music interaction; however, with this definition, it is hard to understand scenarios with loops, for instance, when an "instance" of causes an "instance" of , but then such an "instance" of causes another "instance" of in the future. What does this relation means? Are we traveling back to the time when was first executed? Are we creating a new "instance" of and executing it in the future? Are those two "instances" two different events with the same type (or action)?

The problem of "time travel" becomes even more difficult when we consider multiple instances of a temporal object that could be executed simultaneously. We must distinguish between the motive being repeated and the loop itself; we illustrate some cases in Figure 6. The problem gets even harder when we want to synchronize the ending times of motives and loops. In interactive music, synchronization of loops and motives has been extensively studied by Berthaut *et al.* [18].

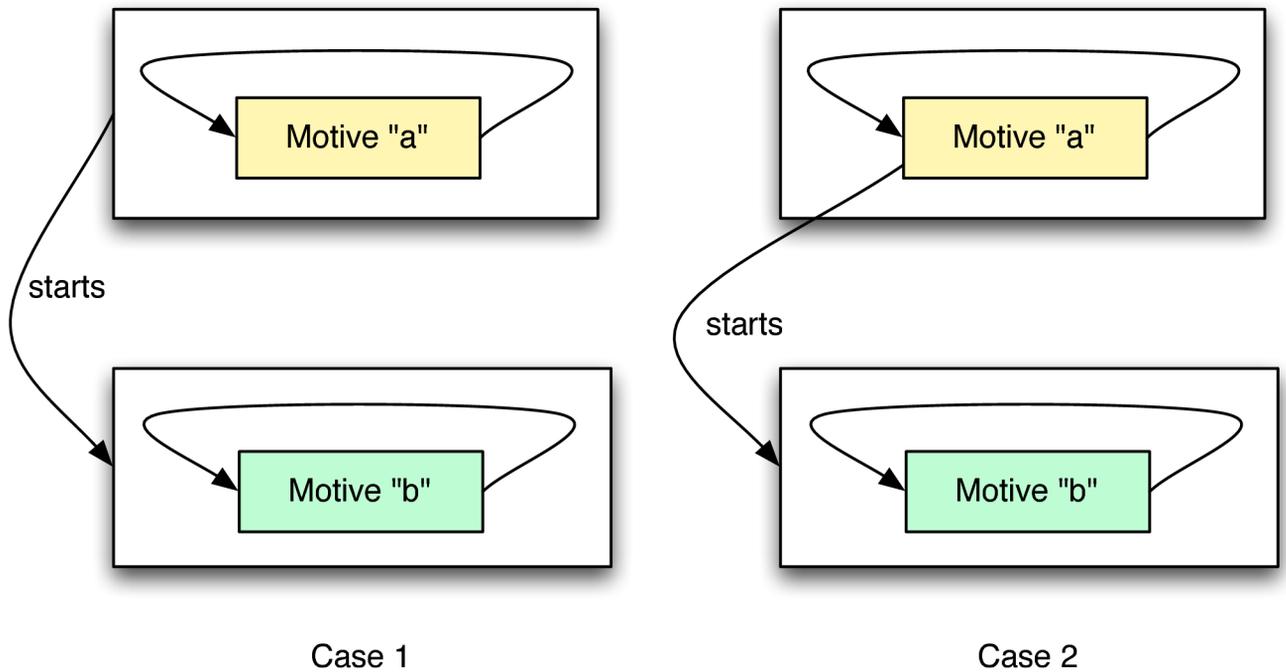

0.52

Case 1        Case 2

Figure 6: Possible scenarios synchronizing motives and loops. In case 1, the loop on the top starts the loop in the bottom; this means that the first repetition of motive "a" starts the first repetition of motive "b". In case 2, every repetition of motive "a" starts a new instance of the loop on the bottom. In case 3, each repetition of motive "a" starts at the same time than each repetition of motive "b".

There are some insights in metaphysics on how to solve the problem of having multiple instances of events. Laudisa argues that in *neoempirism*, leaded by Hume, everything that starts to exist must have a cause for its existence, but all human laws admit exceptions [46]. To formalize such a principle, Laudisa proposes to distinguish between *singular events* and *event classes*: Let $x$ and $y$ be singular events, the existence of a causal connection means that (1) there are event classes of type $X$ and of type $Y$, and (2) $x$ is of type $X$ and $y$ is of type $Y$.

According to Laudisa's postulates, we could think about the start event of a temporal object as a class, and each time the temporal object starts, a different singular event that belongs to such a class is launched. Nonetheless, there is still a problem: how to model choices through time, should we consider a *branching time* or a *linear time*? Let us analyze what computer scientists have to say on this dichotomy.

According to Pratt, there is an analogy: branching time represents local time, and linear time represents global time, in the same way as true concurrency represents local information and false (or interleaving) concurrency represents global information [67]. In linear time, all choices are made at the beginning, it means that we cannot distinguish between a systems that performs actions $a.b+a.c$ from a system that performs $a.(b+c)$, where "." represents sequential composition and "+" represents blind choice. The first system chooses either to execute event $a$ and then event $b$ or event $a$ and then event $c$, whereas the second system executes $a$ and then chooses to execute either $b$ or $c$.

As an example, Vardi argues that with *computational time logic* (CTL), it is possible to characterize bisimulation of concurrent systems. In terms of complexity of the model-checking problem, using CLT is exponentially easier than linear-time logic LTL, but in a competitive analysis, with formulae that can be expressed in both logics, model checkers behave similarly. There is an advantage of linear time: LTL is more intuitive to describe properties because its specifications describe computations, whereas CTL's specifications describe trees of computations [114].

Although branching time seams more appropriate to represent conditional branching in interactive music, we believe that linear time is enough because we can consider that all the temporal objects in a scenario are always executed, but some execute silent actions and some execute visible actions, allowing us to express choices. We want to keep the specification of properties simple.

After analyzing the philosophical problems, the Electroacoustic and Experimental music pieces described above, and existing tools and formalisms for music scenarios, we have identified seven problems with existing software to design music scenario: (1) there is no formal model for music interaction, (2) music scenarios have limited reusability and difficulties with their persistence, (3) time models are temporally unrelated, (4) music interaction software products provide no hierarchy, (5) the different time scales are unrelated, (6) schedulers for music scenario tools are not appropriate for soft real-time, and (7) there is no model to combine temporal relations and conditional branching. In what follows we explain each of those problems.

### There is no formal model for music interaction.

As we explained before, designers usually create music content for their scenarios, and then bind them to external interactive events controlled by *Max/MSP* programs. We advocate a model that encompasses facilities (1) to design music scenarios having complex temporal relationships among components and (2) to define effective mechanisms for synthesis control based on human gestures. We claim that no such model has been proposed.

Such a general model must have formal semantics, as required for automated verification of properties of the scenario that are fundamental to its designers and users. As an example, to verify that temporal objects will be played as expected during performance. In general, we need to prove some property of each execution trace; for instance, that the music motive with notes C-D-E appears in all the traces of execution (or at least in one). Another example is to state that there is at most one temporal object being executed simultaneously. This property is useful in some theater performances to state that there is at most one curtain being moved at the time because of power constraints. Such properties cannot be verified in applications based on informal specifications, as it is the case for most existing scenarios with interactive controls.

### Limited reusability and difficult preservation.

Limited reusability is also a problem caused by the lack of formal semantics: A module made for one scenario might not work for another one because the program may have dependencies on external parameters that are not stated explicitly. The lack of semantics also makes it difficult to preserve music scenarios because there is usually not a score nor a technology-independent precise way for describing the objects, and the temporal and dataflow relations among them.

### Time models are unrelated.

Software to design music scenarios is usually based either on a fixed timeline with a very precise script or a more flexible script using cue lists, as we stated before. A commonly used software to design such scenarios is Live because it allows to use both the fixed timeline and the cue lists, but the two time models are unrelated temporally. In fact, most software products, for instance sequencers, provide only one time model or they are unrelated temporally, as we argued previously.

### No hierarchy for temporal objects.

Most software do not provide a hierarchy to represent the temporal objects of the scenario. As an example, using a hierarchy, it is possible to control the start or end of an object by controlling those from its parent. In interactive music, Vickery argues that using a hierarchy is useful to control higher-order parameters of the piece; for instance, to control the volume dynamics instead of the volume of each note [116]. Concentrating on foreground parameters can lead to music that is too superficial as multiple serialism, according to Vickery.

### Time scales are unrelated temporally.

The different time scales are often unrelated and cannot be controlled in the same tool. *Discrete user gestures* (e.g., clicking the mouse), *control events* (e.g., control messages) and *sound processing* have different sampling frequencies and different computing models. As a consequence of having the time scales unrelated, it is difficult to associate, for instance, a human gesture to both control events and sound processing parameters.

### Event schedules are not appropriate for real-time.

Schedulers for asynchronous dataflow languages (e.g., those from Pd and Max) control both signals and control messages together and they do not support parallelism, thus they often fail to deliver control messages at the required time; for instance, when they work under a high CPU load, which is common when they process video and 3D graphics in addition to sound.

To solve the problem of scheduling and to write high-performance *digital signal processors (DSPs)* for Max and Pd, users often write C++ plugins to model DSPs with loops or independent threads. C++ plugins solve part of the problem, but the control messages –for the input and output of these plugins– are still being scheduled by Max or Pd's schedulers.

Another solution for the scheduler problem –often used during live performance– is to open one or two instances of Max/MSP or Pd simultaneously, running different programs on each one. Nonetheless, synchronization is usually done either manually during performance or by using *open sound control (OSC)*, which adds more complexity and latency.

### No model for conditional branching and temporal relations.

Up to our knowledge, there is not a model for interactive music to represent scores in which is possible to combine complex temporal relations and conditional branching based on conditions over the variables defined in the scenario. In fact, Allombert proposes in [2] an extension with conditional branching to interactive scores, but in such a model he only considers conditional branching and no temporal relations.

## 4 Interactive Scores

In interactive scores, it is possible to specify a variety of relations among temporal objects such as temporal relations, hierarchical relations, harmonic relations, rhythmical constraints and conditional branching. Nonetheless, in this section, we only take into account relations limited to point-to-point temporal relations without disjunction nor inequality ($\neq$) and quantitative temporal relations. We combine qualitative and quantitative temporal relations on the lines of previous independent works by Meiri and Gennary [51, 34].

In what follows, we introduce a mathematic definition of the structure of interactive scores, a formal semantics based on timed event structures, the temporal constraints of a score, and some formal properties such as playability. We also discuss the complexity of the playability problem.

## 4.1 History of interactive scores

The idea of temporal relations among temporal objects was introduced by Beurivé and Desainte-Catherine in [19]. They found out that relative times are a more efficient representation than absolute times for music scores. Soon after, they developed *Boxes*: a software to model a hierarchy and temporal constraints [19]. In fact, Boxes uses Allen's relations to describe temporal constraints. A few years later, Desainte-Catherine and Brousse came up with the idea of the interactive scores formalism [29].

Another system dealing with a hierarchy of temporal objects is *Maquettes of OpenMusic* [21]. However, we argue that *OpenMusic* [22] is a software for composition and is not meant for real-time interaction. Allombert and Desainte-Catherine figured out that the music interaction community needed a software for composition capable of describing a hierarchy of temporal objects and capable of real-time interaction! In 2005, they introduced a new model of interactive scores [7], extending the previous model developed by Desainte-Catherine and Brousse, and following the concepts of Haury's meta-instrument [39]. This model admits modification of the starting and ending times of the notes of the score during execution.

In Allombert and Desainte-Catherine's new model, a score is composed by temporal objects, interactive events and temporal relations. This approach does not allow to define interactive user events inside the hierarchy, as we can do it today. They extended Allen's relations with quantitative relations to express the duration of temporal objects in a similar manner as Mieri did it back in 1995. They introduced the very first notions of temporal reduction: intervals can be reduced if an event is launched before its nominal (expected) time and intervals can be extended if the event is launched after its nominal time; however, the operational semantics of the temporal objects with nominal times, was not very well defined back then. They also introduced a semantics based on Petri nets. Finally, they introduced the *environment, control, output (ECO)* machine: an abstract machine to execute an interactive score in real-time.

Allombert, Desainte-Catherine and Assayag presented a new extension in 2007 [3]. They changed the definition of a score: A score is defined as a pair $\langle T,R \rangle$ where $T$ is a set of temporal objects and $R$ a set of temporal constraints. This new definition considers an interactive user event as a kind of temporal object, thus they are included in the hierarchy, as opposed to the extension they presented in 2005. They also argued that interactive scores must have two modes: the edition mode, which they implemented using constraint propagation, and the execution mode, which they made using Petri nets. The edition model is a linear constraint satisfaction problem with a cyclic constraint graph, according to Allombert *et al*.

In the extension of interactive scores developed in 2007, Allombert *et al.* realized that some transformations were needed to the Petri nets to execute them properly. They proposed to collapse two places that occur at the same time in the same place (state). Those transformations inspired what we call in this dissertation the *normal form*. They also introduced global constraints, but not the details on how to implement them. They also developed an implementation using OpenMusic. The implementation they made in OpenMusic will be the base for the *Iscore* library, developed one year after.

In 2008, Allombert *et al.* developed a new extension of interactive scores [9]. They introduced a new kind of temporal relations: linear constraints over durations; for instance, to say that the duration of an object is $k$ times bigger than another. They made an implementation in OpenMusic that only includes flexible-time durations and does not include linear constraints. Examples of their quantitative relations are those involving a proportional or explicit duration; for instance, "the duration of *A* is one third of the duration of *B*" or "the duration of *A* is 3 seconds". Examples of their qualitative temporal relations are those to represent the precedence between the start and end points of two temporal objects; for instance, "*A* must be played during *B*" or "*C* must be played after *D*". They also improved the concept of *temporal reductions*: left reductions (chronological) and right reductions (anti-chronological). Temporal reductions are a mechanism to reduce or stretch the duration of a temporal object when an interactive event is, respectively, delayed or speeded up, while respecting the temporal constraints of the score.

It was most likely that they realized at that time that including linear constraints over the duration of the temporal objects will change the complexity of the satisfiability and dispatching of the temporal constraints because they could no longer represent the temporal constraints as a simple temporal problem. Constraints over the durations of temporal objects were never again presented in interactive scores models.

Allombert *et al.* explored other alternatives to Petri nets as semantics for interactive scores. After reading all the previous extensions of interactive scores, Rueda had in mind that a process calculus based on constraint programming would be more appropriate to represent temporal constraints (and even other constraints, such as harmonic and rhythmical) than Petri nets. Rueda worked with Allombert, Assayag and Desainte-Catherine to develop a model based on ntcc in 2006 [5]. They used Allen's relations as temporal relations. There is a disadvantage: The model does not consider the problems that arises when two objects are constraint to start at the same time nor the problems associated to dispatching efficiently a simple temporal problem, as described by Muscettola *et al.* [55].

Sarria found another disadvantage with the ntcc model of interactive scores developed by Rueda *et al.*: time units in ntcc may have different (unpredictable) durations. Sarria extended Allombert's model in his Ph.D thesis in 2008. He proposed a different approach to cope with real-time issues using his own CCP variant, *the real-time concurrent constraint (rtcc)* calculus [85]. Rtcc is an extension of ntcc capable of modeling time units with fixed duration. This new calculus is capable of interrupting a process when a constraint can be inferred from the store. Rtcc is also capable of delays within a single time unit.

Olarte *et al.* also extended Rueda's ntcc model. They extended the model to change the hierarchy of temporal objects during execution [58]. The spirit of such a model is different: they focus on changing the structure of the score during execution to allow the user to "improvise" on a written piece, whereas we are interested on a simpler model that we can execute in real-time. It is worth noticing that it may be also possible to model such changes in the structure during execution using a special kind of Petri nets in which tokens are also nets, introduced by Köhler *et al.* [43].

Finally, in 2009, Allombert explained in his Ph.D. the results published previously in his models [2]. He also introduced some ideas on how to deal with durations of arbitrary intervals, he introduced music processes that can be associated to temporal objects, and he introduced conditional branching. Conditional branching is the base for some non-linear models in music. Non-linear models are used to create openworks. Open works can have openness of interpretation or openness of semantic content, as explained by Vickery [115].

Allombert presented in his thesis conditional branching and temporal relations separately, but he did not show an unified way to represent conditional branching together with temporal relations in the same scenario. His work on conditional branching was partially based on previous results developed during Ranaivoson's M.Sc. thesis in 2009 [71]. These two works are the base of our conditional branching extension.

## 4.2 Interactive Scores Formalism

There are formalisms to model interactive scenarios such as *interactive scores*. Interactive scores has been a subject of study since the beginning of the century [29]. The first tool for interactive scores is *Boxes*, developed by Beurivé [19]. Boxes was conceived for the composition of Electroacoustic music with temporal relations; however, user interaction was not provided. A recent model of interactive scores [2], that significantly improves user interaction, has inspired two applications: *i-score* [4] to compose and perform Electroacoustic music and *Virage* [6] to control live performances and interactive exhibitions. We give a further discussion on the history of interactive scores. Scenarios in interactive scores are represented by *temporal objects*, *temporal relations* and *interactive objects*. Examples of temporal objects are sounds, videos and light controls. Temporal objects can be triggered by interactive objects (usually launched by the user) and several temporal objects can be executed simultaneously. A temporal object may contain other temporal objects: this hierarchy allows us to control the start or end of a temporal object by controlling the start or end of its parent. Hierarchy is ever-present in all kinds of music: Music pieces are often hierarchized by movements, parts, motives, measures, among other segmentations.

Temporal relations provide a partial order for the execution of the temporal objects; for instance, temporal relations can be used to express precedence between two objects. As an example of *relative temporal relations*, the designer can specify that a video is played strictly before a light show or between 10 and 15 seconds before. As an example of *absolute temporal relations*, the designer can specify that a loop starts three seconds after the video.

### New semantics for interactive scores.

We provide an abstract semantics for interactive scores based on *timed event structures*. The purpose of such a semantics is (1) to provide an easy, declarative way, to understand the behavior of a score, and (2) a simple theoretical background to specify properties of the system. In constraint programming, we can specify some properties of the scores such as playability. We can also specify those properties in event structures; moreover, the notion of *trace*, inherent in event structures, is more appropriate than temporal constraints for certain properties. As an example, to specify that a music motive appears in at least one trace of execution.

This study led us to discover that there is no difference between interactive objects and the other temporal objects in the event structures semantics: such a difference can only be observed in the operational semantics. That was the main reason to introduce an operational semantics based on ntcc, on the lines of Allombert *et al.* [5]. Nonetheless, in Allombert *et al.*'s models of interactive scores, it was not precisely stated how to execute scores whose temporal object durations are arbitrary integers intervals; for instance, a score in which object *a* must be executed between two and four time units after object *b*. Allombert *et al.*'s models handle *flexible-time intervals*: {0} to express simultaneity, and (0,∞) and [0,∞) for precedence or for the flexible duration of the objects. Allombert *et al.*'s models also miss an abstract semantics.

We extend the interactive scores formalism with an abstract semantics based on event structures and an operational semantics specified in ntcc, providing (1) a new insight into the interactive scores model; (2) more complex temporal relations to bind objects, including arbitrary sets of integers in the event structures semantics and arbitrary intervals in the operational semantics; and (3) the possibility to verify properties over the execution traces. In order to use arbitrary integer intervals in our operational semantics, we show that several transformations to the event structures semantics are needed to define operational semantics that can dispatch the temporal objects of the score in real-time.

To complete our framework, we also present in this dissertation two extensions of the interactive scores formalism: one for conditional branching and one for signal processing. We also explain the implementation of interactive scores and the implementation of an automatic verification tool for ntcc.

### Time conditional branching interactive scores.

Non-linear music pieces are *open works*. According to Vickery, open works may have openness of interpretation or openness of semantic content [115]. Conditional branching is essential to describe pieces with openness of interpretation.

Conditional branching is commonly used in programming to describe control structures such as *if/else* and *switch/case*. It provides a mechanism to choose the state of a program based on a condition and its current state. In music interaction, using conditional branching, a designer can create scenarios with loops and choices (as in programming).

In the domain of interactive scores, using conditional branching, the user or the system can take decisions on the performance of the scenario with the degree of freedom that the designer described. The designer can express under which conditions a loop ends; for instance, when the user changes the value of a certain variable, the loop stops; or the system non-deterministically chooses to stop. As an example, the designer can specify a condition to end a loop: When the user changes the value of the variable *end* to `true`, the loop stops. The designer can also specify that such choice is made by the system: The system non-deterministically chooses to stop or continue the loop.

We chose event structures because it is a powerful formalism for concurrency that allow us to extend the interactive scores semantics with conditional branching and loops in a very precise and declarative way. Conditional-branching timed interactive scores were introduced in [102, 101]. Such an extension has operational semantics based on ntcc, but it misses an abstract semantics to understand the conflicts among the temporal objects that take place when modeling conditions and choices.

## Interactive scores with signal processing.

It is crucial that interactive music software products preserve the *macroform* and the *microform* of the scenario. The macroform includes the structure of the scenario (e.g., the tempo and the duration of the scenes, movements, parts and measures). The microform comprises the operations with samples (e.g., micro delays, articulation, intonation, and envelop of the sound).

We propose an extension to the interactive scores formalism for sound synthesis. In this extension, we deal with microstructure and macrostructure of sound, not the structure of image nor other media. In the interactive scenarios we consider, we can deal with streams produced in real-time (e.g., a stream captured from the microphone).

We define a new type of temporal relations meant for high precision; for instance, to express micro delays. We also introduce *dataflow relations*; for instance, how the audio recorded by a temporal object is transferred to another object to filter it, add a micro delay, and then, send it to another temporal object to be diffused.

We also propose an encoding of the scenario into two models that interact during performance: (1) A model based on the ntcc for concurrency, user interactions and temporal relations, and (2) a model based on Faust for sound processing and micro controls. An advantage of having a formal model for ntcc and Faust interoperation is that we could prove properties such as playability, and predict the behavior of the system.

The novelty of our approach is using the constraints sent from ntcc to control Faust. We tested our examples in Pd, although they could also be compiled for Max or as a standalone program since both Faust and ntcc can be translated into C++ and Max. In fact, the final goal of our research is to develop a standalone program for interactive scores. Such a program should be general enough to interact with Pure Data, Live, Max/MSP and other existing software either by passing messages or by generating plugins for those languages.

## Execution of interactive scores.

We give operational semantics for interactive scores, but we need to execute those models. The execution must be able to interact with a user in real-time. Since the operational semantics are given in ntcc, we need an interpreter for ntcc capable of real-time interaction and being able to control music objects such as sound, video and lights.

There are some interpreters for ntcc, but they are not suitable for real-time interaction [54, 75]. We chose a real-time capable interpreter for ntcc, *Ntccrt* [100], to execute our models. Ntccrt is based on *Gecode* [90]: state-of-the-art in constraint propagation. Ntccrt programs can be compiled into standalone programs, or plugins for Pd or Max. Users can use Pd to communicate any object with the Ntccrt plugin. In fact, Ntccrt can control all the available objects for audio processing defined in Pd, although our goal is to use Faust for such tasks.

Ntcc belongs to a bigger family of process calculi called *concurrent constraint programming* (CCP). In the last decade, there has been a growing interest for CCP models of music interaction [77, 80, 78, 79, 75, 5, 100, 58, 59, 105][12].

Ntcc is not only useful for music semantic interaction, ntcc has also been used in other fields such as modeling molecular biology [76], analyzing biological systems [36] and security protocols [47]. Therefore, advances on the simulation of ntcc models will be useful not only for music interaction, but also for other fields.

**Automatic verification.**

A disadvantage of ntcc is the lack of automatic verification tools available. This limits the applicability of the verification techniques to small problems. We claim for the urgent need of a verification tool for ntcc. First, because ntcc has been widely used to model reactive systems and verify properties about them, but the verification had to be done by hand. Second, because there are not many frameworks to model and verify music interaction systems, and ntcc has been proved to be successful in that field.

We developed a bounded-time model checking procedure for ntcc, *ntccMC*[13]. The model checker is based on encoding ntcc processes and *constraint linear-time logic* ( CLTL) formulae into deterministic finite state automata. Examples of CLTL formulae are "always the constraint *pitch*=60 can be deduced from the output store", namely (*pitch*=60); and "eventually object *a* and object *b* are launched at the same time", namely .

Ntcc has been used since its beginnings to prove properties of music interaction systems. Ntcc is a powerful formalism because it allows to simulate the behavior of a model and also to verify properties of the model. As an example, ntcc was used to verify properties of a musicological problem of western-african music [77]. The reader may also look at [78] and [80] for other examples of verification of music interaction systems.

## 4.3 Structural Definition of the Score

Interactive scores are composed by temporal objects and temporal relations. We consider that all temporal objects have only a start and end point and it is not possible to define intermediate points.

### 4.3.1 Temporal objects

A temporal object has two *point identifiers*: to control its starting and ending times. An external action is usually associated to each of them (e.g., turn on the lights, play a video or stop a sound). Some temporal objects are interactive, thus we call them *interactive objects*.

**Definition 1 (Temporal object (TO) )** *Let P be a set of point identifiers. A* Temporal object *is a tuple o=⟨sp,ep,Δ⟩, where sp,ep∈P,sp≠ep, are called* start *and* end *points, respectively, and Δ⊆N is a set of* durations. *A temporal object whose duration Δ={0} is called an* interactive object. *Functions sp(o), ep(o) and d(o) return the start, end and duration, respectively, of object o. The set of all temporal objects is* 𝒯

---

### 4.3.2 Temporal relations

Points $p,q \in P$ are supposed to be positioned on a timeline. Temporal positions of points could be fully or partially determined. Temporal relations constrain the set of possibilities for these positions. A partial order among points is given by *quantitative relations*; for instance, point $q$ is executed between four and ten time units after point $p$. *Qualitative temporal relations* can be easily expressed as quantitative relations; for instance, point-to-point *before* relation is the interval $(0,\infty)$ and point-to-point *equal* relation is the set $\{0\}$, a proposed in [51].

Our quantitative relations are close in spirit to the temporal relations described by Allombert *et al.* which contain time intervals [9]. A limitation of Allombert's interactive scores is that all intervals must be *flexible*: intervals must have the form $(0,\infty)$, $[0,\infty)$ or $\{0\}$. In Allombert's thesis [2], the model is extended to general integer intervals, but arbitrary durations cannot be expressed. The durations contained in our temporal relations are usually intervals, but they can be any set of integers.

**Definition 2 (Temporal Relation)** *Let function give the set of potential time positions for each point* $p \in P$. *A temporal relation is a tuple* $\langle p, \Delta, q \rangle$ *where* $\Delta \subseteq N$ *is the set of durations between points* $p,q \in P$. *We use the notation for temporal constraints of duration. Temporal positions of p and q are said to be constrained by* $v(q) = v(p) + \Delta$. *The set of all temporal relations is* $\mathcal{R}$.

*Allen's relations* [1] without disjunction, over discrete time, can be easily expressed as point-to-point relations [51]. Furthermore, with point-to-point relations we can express relations that cannot be expressed in Allen's relations without disjunction; for instance, that the end of a temporal objects is before the end of another temporal object.

**Example 3** *Figure 7 shows how the Allen's relation "red light overlaps green light" can be represented by three point-to-point before relations.*

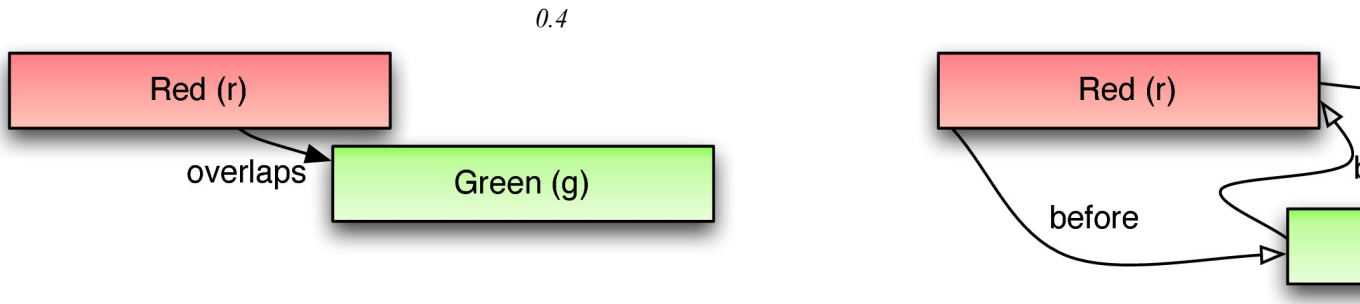

*Figure 7: Encoding of the Allen relation overlaps into point-to-point relations.*

### 4.3.3 Interactive scores

**Definition 4 (Interactive Score)** *An* interactive score *is a set of temporal objects equipped with a set of temporal relations: a tuple $\langle P,O,R \rangle$, where $P$ is a set of point identifiers, $O \subseteq \mathcal{T}$ is a set of temporal objects, $R \subseteq (P \times P(N) \times P)$ the temporal relations. Set $R$ also includes the relations derived from the duration of temporal objects. For each $o \in O$, $\langle sp(o), d(o), ep(o) \rangle \in R$. In addition, a relation $\langle p, \Delta, q \rangle \in R$ iff*

1. *$p,q$ are distinct points and $v(q)=v(p)+\Delta$;*

2. *two interactive objects do not occur at the same time; and*

3. *there is only one temporal relation between the start and end point of a temporal object.*

Property 2 takes care of the fact that two interactive points cannot happen at the same time; it means, that they cannot be related with zero-duration temporal relations, not even transitively by the means of other objects. The reason for this constraint is that interactive objects are usually launched by the user of the scenario; therefore, we cannot guarantee that the user will launch them at the same time. This simplifies the model.

**Example 5** *Figure 8 is an example of a score. Objects red light, green light and sound produce visible actions at their start and end. Objects a,b are interactive. Temporal relations starts represents a zero-duration between the start points of the two objects they connect. Relations ends represents a zero-duration between the end points of the two objects they connect. Allen's relation overlaps can be represented by the three point-to-point relations, as shown in Figure 7.*

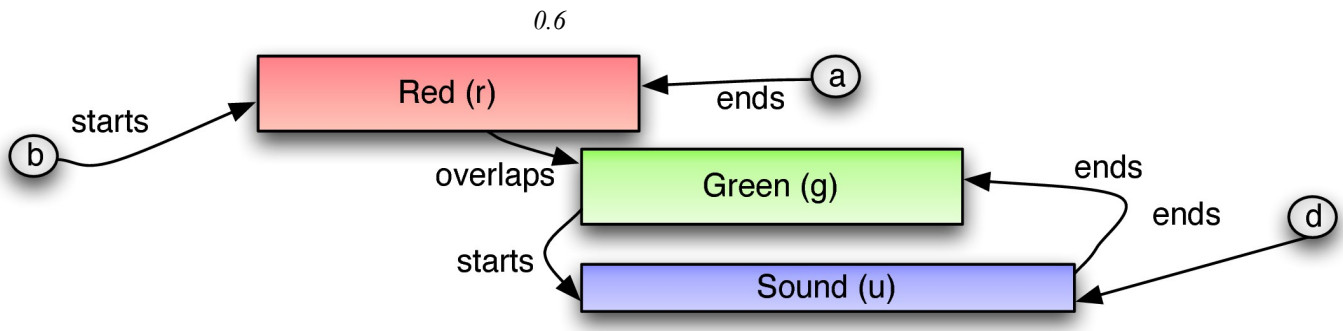

*Figure 8: Example of an interactive score.*

**Example 6** *The constraints of the score in Figure 8 are presented in Table 3.*

| Constraints of duration | Explicit temporal relations |
|---|---|
| $\nu(ep(r))\in\nu(sp(r))+d(r)$ | $\nu(sp(g))=\nu(sp(u))$ |
| $\nu(ep(g))\in\nu(sp(g))+d(g)$ | $\nu(ep(a))=\nu(ep(r))$ |
| $\nu(ep(a))\in\nu(sp(a))+\{0\}$ | $\nu(sp(g))>\nu(sp(r))$ |
| $\nu(ep(b))\in\nu(sp(b))+\{0\}$ | $\nu(ep(g))>\nu(ep(r))$ |
| $\nu(ep(d))\in\nu(sp(d))+\{0\}$ | $\nu(sp(g))<\nu(ep(r))$ |
| $\nu(ep(u))\in\nu(sp(u))+d(u)$ | $\nu(sp(d))=\nu(ep(u))$ |
| | $\nu(sp(b))=\nu(sp(r))$ |

Table 3: Implicit and explicit temporal constraints of the score in Figure 8. Relations "<" and ">" are represented by the interval $(0,\infty)$; relation "=" is represented by the set $\{0\}$. 8

## 4.4  Event Structures Semantics

Langerak's *timed event structures* (henceforth *event structures*) is a mathematical model to represent systems with non-determinism, real-time and concurrency [13]. Event structures allow to define a partial order among concurrent events. Event structures include a set of *labeled events* and a *bundle delay relation*. The bundle delay relation establishes which events must happen before some other occurs. Actions can be associated to events. Events are unique, but two events may perform the same action. Events can be defined to be "urgent". An *urgent event* occurs as soon as it is enabled. In addition to the bundle relation, event structures include a *conflict relation* establishing events that cannot occur together. Events can also be given *absolute occurrence times*.

We recall that interactive scores must have formal semantics, as required for automated verification of properties of the scenario that are fundamental to its designers and users. We also recall that we denote by the functions and $R(\varepsilon)$ each component of an event structure $\varepsilon$.

### 4.4.1  Temporal objects

The events represent the start or end points of a temporal object. An interactive object is represented by a single event. Temporal relations are modeled with event delays. A static temporal object $a$ is represented by two events  (start and end events). The labels of events are pairs (*type*,*o*), where *type*$\in\{startPoint,endPoint,interactiveObject\}$ and *o* is the temporal object giving rise to the event.

**Example 7** *Figure 9 shows the encoding of three temporal objects.*

**Definition 8 (Temporal object encoding)**  *The encoding of a temporal object (a) is a function  defined by*

1. *if a=⟨sp,ep,{0}⟩ (i.e., a is an interactive object),*

   *then*

2. *if a=⟨sp,ep,Δ⟩ (i.e., a is a static temporal object), then eto(a)=⟨E,l,∅,⟩,*

*where  and .*

The above definition guarantees that there are unique start and end events in the translation of a static temporal object, thus we know that each event is related to a single point.

**Definition 9 (Relation between points and events)**  *Let o be a temporal and  the set of points contained in o, function  associates a point identifier  to its corresponding event in eto(o).*

0.5

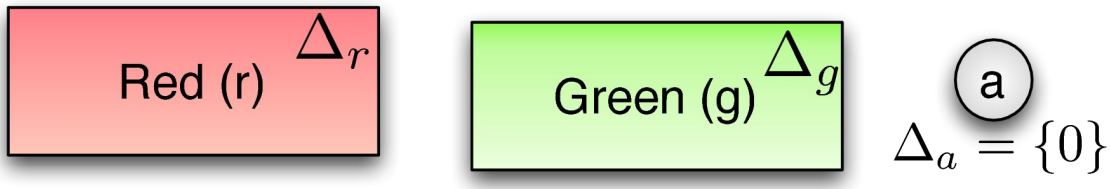

Figure 9: Encoding of a temporal object and its temporal relations of duration. There are two for *r*, two for *g*, and a single one for *a*. Double line arrows are just a visual notation for the event delays that model the duration of the temporal objects.

### 4.4.2  Temporal relations

Each point-to-point relation is represented by an event delay function.

**Definition 10 (Temporal relation encoding)**  *Let* p *be a point of temporal object* a *and* q *be a point of temporal object* b. *The encoding of a temporal relation r is given by the function . For each r=⟨p,Δ,q⟩∈ℛ the encoding etr(r) is defined by  pe(b,q).*

**Example 11** *Figure 10 is the encoding of an overlaps relation between the objects r and g.*

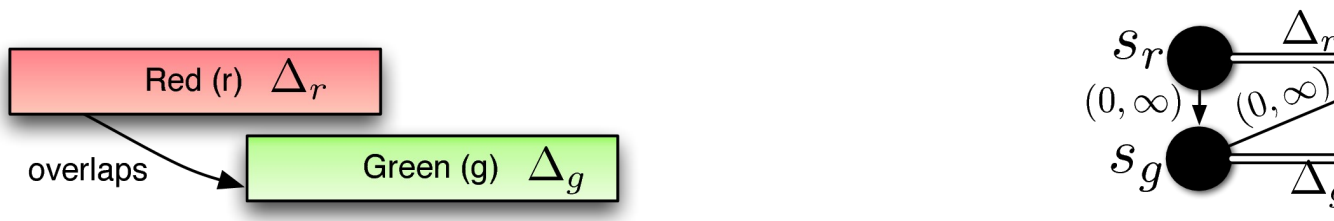

*Figure 10: Encoding of two temporal objects, and the overlaps relation between them.*

### 4.4.3 Interactive scores

The encoding of a score is given by adding the event delays from the encoding of the temporal relations to the encoding the temporal objects.

**Example 12** *The encoding of Figure 8 is presented in Figure 11.*

**Definition 13 (Interactive score encoding)** *The encoding of an interactive score $s=\langle P,O,R\rangle$ is given by the function that translates interactive scores into event structures. Let , then .*

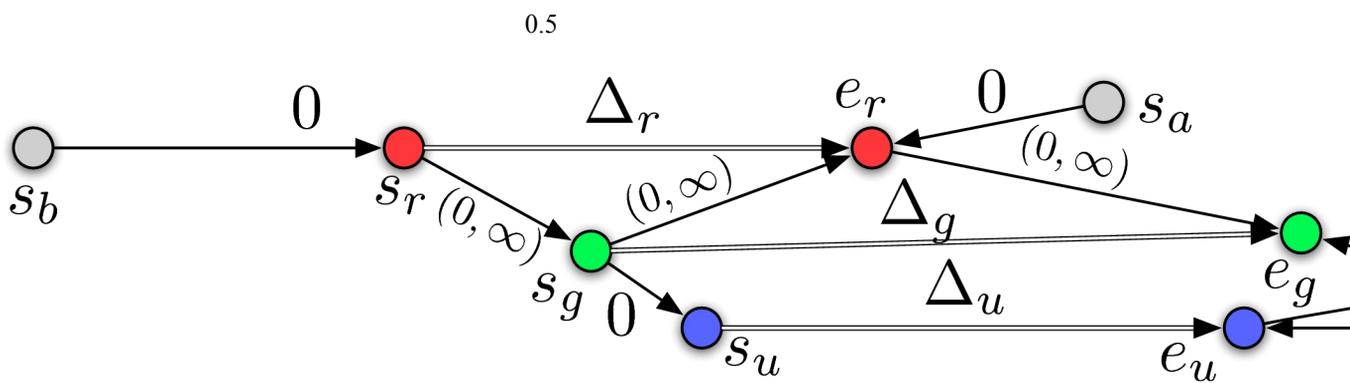

Figure 11: Encoding of the score in Figure 8.8

We shall prove that the temporal constraint of the event structures semantics of a score corresponds to the temporal constraint of the score.

**Definition 14 (Temporal constraint of an event structure)**   *Let ε=⟨E,R,l⟩ be an event structure without conflicts. The temporal constraint of an event structure tc(ε) is the conjunction of constraints  for each  with , where  is a finite set of natural numbers.*

Given an event structure ε,  is a valid trace of ε iff  is a solution to *tc(ε)*. The proof proceeds as follows. By the definition of event structures without conflicts Error: Reference source not found, for all 0<*i,j*≤*n*:  in any trace of ε because ε has no conflicts. By Def. , for each , we have the constraint . Therefore,  is a solution to *tc(ε)*.

**Proposition 15 (Equivalence of interactive score constraints and  its event traces)**   *Let s=⟨P,O,R⟩ be an interactive score, ε=es(s) the encoding of the score, ts(s) the temporal constraint of the score, and tc(ε) the temporal constraint of ε. It holds that , where  is obtained by replacing each point identifier by its corresponding event in the constraint ts(s), and p is the start or end point of temporal object c∈O.*

We recall that  gives the set of *potential time positions* for each point *p*∈*P*. We also recall the notation for temporal constraints: *t*+Δ={*t'*|*t'*=*t*+δ,δ∈Δ}.

<center>[Sorry. Ignored \begin{proof} ... \end{proof}]</center>

The proof above is presented for hierarchical interactive scores in [105].

# 4.5  Some Properties of the Scenarios

We insist that a motivation of defining an abstract semantics in event structures is to prove properties of the system execution; in particular, properties about the execution traces. As an example, to verify that temporal objects will be played as expected during performance (i.e., *playability*) or, in general, some property of each execution trace. Such properties cannot be verified in applications based upon informal specifications, as it is the case for most existing software for music scenarios with interactive controls. The following properties were already presented in [105].

- **Properties of the traces of execution.**

  - There exist a trace σ that contains a word *w*; for instance, the sequence of notes C-D-E is part of *n* traces of execution.

  - There exists *n* traces σ that contain a word *w*, possibly with other events in between; for instance, the sequence of notes C-D-E is contained in the trace .

  - The number of possible traces of execution for a score ε is *card*(*Traces*(ε)).

  - If event *e* is launched before time unit *n*, the duration of object *a* is greater than *m*. For all σ∈*Traces*(ε) and , it holds that .

  - After event *e* is played, there are *n* traces where event *f* is launched before event *g*.

  - Between time units *a* to *b*, there is no more than *n* objects playing simultaneously.

- **Minimum duration of a score.** Let *s* be a score and ε=*es*(*s*) the encoding of *s*, the trace whose duration is minimum corresponds to a path from the start event of ε to the end event of ε such that the sum of the delays in the event delay relation is minimal among all paths connecting start and end.

- **Maximum and minimum number of simultaneous temporal objects.** Let be a trace of $\varepsilon=es(s)$, and $maxS(\sigma)$,$minS(\sigma)$ the maximum and minimum number of events executed simultaneously in $\sigma$, respectively. The maximum and minimum number of simultaneous temporal objects of a score correspond, respectively, to the maximum and minimum value of $maxS(\sigma)$ and $minS(\sigma)$ among all $\sigma \in Traces(\varepsilon)$. This property is useful, for instance, to argue that there is only one curtain moving at the time during a theater performance.

- **Playability of a score.** This property states that all temporal objects will be *played* during execution; this is desirable because a score can be over-constrained and therefore not playable. Formally, let be the events played in trace $\sigma$. We say that a score is *playable iff* for all $\sigma \in Traces(es(s))$ it holds that .

    The playability of a score can be decided by solving a *constraint satisfaction problem (CSP)*. There exists a $\sigma \in Traces(es(s))$ such that *iff* the following CSP has at least one solution: a variable for each event ; the domain for each variable, where is a finite subset of $N$; and the single constraint $tc(\varepsilon)$. This holds as a direct consequence of Prop. .

### 4.5.1 Time complexity of the playability of a score

In what follows we will show that deciding the playability of a score is NP-complete in the general case, but there is an interesting subclass that is tractable.

#### The NP-complete case

We will show that the decision problem of the *subset sum* [50] can be encoded as the playability of an interactive score. The subset decision problem is stated as follows: Given a set of integers of $n$ objects and an integer $W$, does any non-empty subset sum to $W$?

$$(1)$$

There are several algorithms to solve the subset sum, all with exponential time complexity in $n$, the number of objects. The most naïve algorithm would be to cycle through all subsets of $1 \leq k \leq n$ numbers and, for every one of them, check if the subset sums to the right number. The running time is of order O(), since there are subsets and, to check each subset, we need to sum at most $n$ elements. The best algorithm known runs in time O(), according to Martello [50]. In what follows we show that the playability of a score is a NP-complete problem by following the methodology described in [87].

**Proposition 16 (The playability of a score is a np-complete problem)**

*(1) The subset sum decision problem can be encoded as the playability of an interactive score. (2) If the score is not playable, there is not a subset whose sum is W. (3) If the score is playable, then it exists at least a subset whose sum is W. (4) To check whether a solution satisfies the playability problem can be done in polynomial time.*

[Sorry. Ignored \begin{proof} ... \end{proof}]

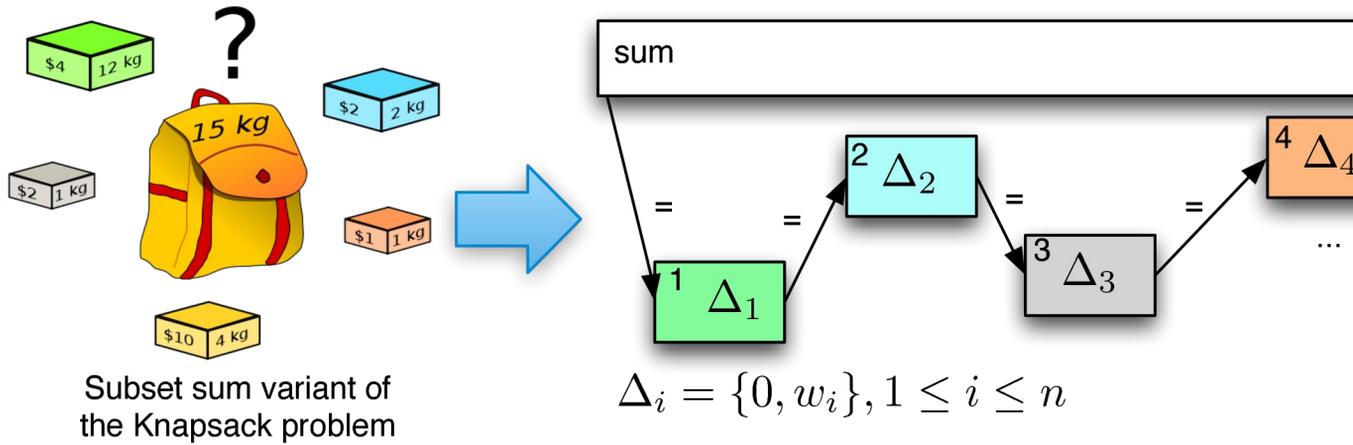

Figure 12: Encoding of the subset sum problem into an interactive score. Note that the subset sum problem is a variant of the knapsack decision problem where costs are not taken into account and the goal is to find if there is a subset of the elements that fills exactly the knapsack capacity.

### 4.5.2 A polynomial-time subclass

The conjunction of temporal constraints of an interactive score can be represented as a *simple temporal problem (STP)* when the domains of the durations are intervals of integers without holes [27]. The translation of the playability of a score into a STP consists in a set of point variables , one for each point in the score, and a set of binary constraints over those variables , one for each temporal constraint of the score. Each constraint has the form  with  and *a,b*∈*N*∪∞. Constraints of the form  can be easily obtained from the temporal constraint of an interactive score, defined in Equation Error: Reference source not found. As an example, constraint of the form  can be translated into two constraint a constraint . It is left to the reader the encoding of the inequalities into constraints of the form .

The satisfiability of a STP can be easily computed with an algorithm to find *all-pairs shortest-path* of a graph, such as *Floyd-Warshall* [24] algorithm which has a polynomial time and space complexity. In fact, Floyd-Warshall has a time complexity of , where *n* is the number of points of the score. There are faster algorithms for this problem in the literature [66, 118]; however, they are efficient to calculate if a STP has a solution, but do not guarantee that the constraint problem remains satisfiable when dispatching the events during the execution of a score.

Fortunately, with some transformations, a STP can be dispatched online efficiently by relying only on local propagation: looking only to the neighbors of the event launched, as proposed by Muscettola *et al.* [55]. We extend the approach of Muscettola *et al.* to event structures: Transform an event structure in such a way that the events of the event structure can be dispatched online efficiently.

**Iscore.**

*Iscore* **is a library developed by Allombert *et al.* that implements the ECO machine to execute interactive scores. It was originally developed in Lisp, and then it was** ported to C++ during the ANR *Virage*[14] project in 2008. Allombert *et al.* introduced Iscore as a new tool that replaces Boxes [4]. The comparison with Boxes is given in detail in [8]. Iscore uses Petri nets as its underlying model because Allombert argued that solving constraint satisfaction problems during execution may be incompatible with real time [2]. The first implementation of Iscore uses the OpenMusic Maquettes environment and the constraint solving library Gecode in the edition mode. During execution, OpenMusic communicates with Max or Pd. Max and Pd are in charge of the contents of the temporal objects. The communication is done using the *open sound control (osc)* protocol. The library was ported to C++ during the project *Virage* and it is currently being used by *Acoumouscribe*.

**Virage.**

Virage is a software that uses Iscore and provides a user-friendly interface for edition and execution of interactive scores [6]. It was designed for interactive theater performances, but it can also be used for Electroacoustic music. Recently, Marczac *et al.* describe an extension of Virage with *fast forward* and *go to jumps* functionalities [49]. *Fast forward* is used to modify the execution speed of the score, and *go to jumps* can be seen as very fast a acceleration in which the artist do not want intermediate values.

**Acousmouscribe.**

"The Acousmoscribe is a free software coming from the former software, *Boxes*, which aim was to write scores and compose electroacoustic music. Acousmoscribe is built around two possible uses: notation and composition. This software offers concrete and symbolic approaches of electroacoustic music at the same time. The user interface allows the writing of electroacoustic music scores, following the phenomenological approach initiated by Pierre Schaeffer. Around twenty signs, that can be combined thanks to a palette to write a "sound object", produce more than 20000 combinations: In this way, its use is intuitive while allowing quite a precise description of sounds. The length of each created box corresponds to the length of the associated sound in time. Regarding composition, a software built in Max/MSP named Acousmosynth receives messages from Acousmoscribe thanks to the open sound control protocol, and translates its symbolic notation into control parameters for audio synthesis modules." [15]

---

14 ANR Virage Project Virage was a research platform project that aimed at developing new writing and management interfaces for artistic creation and cultural industries. This platform included businesses (JazzMutant, Blue Yeti, RSF), academic laboratories (LIMSI-CNRS Paris Sud, MSH Paris Nord-CICM, LaBRI Bordeaux) and artists (GMEA, the Albi-Tarn centre national de création musicale and didascalie.net).

15 http://scrime.labri.fr/index.php?option=com_content\&view=article\ &id=11\%3Aacousmoscribe\&catid=41\%3Athemesderecherche\&Itemid=81\ &lang=en

**i-score.**

The latest software for interactive scores is *i-score*. This software combines the edition interface of Acousmouscribe with the execution model of Virage. It is currently maintained by *Scrime*[16] and distributed as *opensource*.

# 5 Other Software and Formalisms in music interaction

In what follows we describe software and formalisms used in music interaction such as sequencers, signal processing languages, dataflow languages and process calculi.

## 5.1 Sequencers

Software to design music scenarios are usually based either on a fixed timeline with a very precise script or a more flexible script based on cue lists. As an example of fixed-timeline sequencers, there are two well-known sequencers for *Mac OS X*: *Pro tools*[17] and *Final cut pro*[18]. As another example, the theater cue manager *Qlab*[19] is based on cue lists. In Qlab, the user programs a list of upcoming events; however, Pro tools, Final cut pro and Qlab only use one time model and cannot use both.

Another software to design music scenarios is *Ableton live*[20]. Live is often used in Electroacoustic music and performing arts because it allows to use both the fixed timeline and the cue lists. Nonetheless, both time models are unrelated temporally.

An advantage of interactive scores over the previously mentioned sequencers is to relate temporally both time models and to model conditional branching.

## 5.2 Computer-assisted improvisation

*Computer-assisted improvisation* **usually considers building representations of music, either by explicit coding of rules or applying machine learning methods. An interactive machine improvisation system capable of real-time must perform two activities concurrently: stylistic learning and stylistic simulation. As an example, the *Omax* system [12, 48] and the *Continuator* [61] construct models to represent the sequences played by the musician and create their own sequences based on the musician's style.**

---

16 http://scrime.labri.fr/
17 http://www.avid.com/us/products/pro-tools-software
18 http://www.apple.com/finalcutpro/
19 http://figure53.com/qlab/
20 http://www.ableton.com/

Improvisation systems are interactive and concurrent, but they are different to interactive score systems: their goal is to create music based on the user style, whereas interactive scores is a formalism to compose music (or create music scenarios). In interactive scores, the designer describes several rules that have to be respected during execution and the system does not produce new sequences nor sounds that are not written in the score.

## 5.3 Meta-instruments

A meta-instrument is a musician-machine interface and a gesture transducer intended for Electroacoustic music, music work, and, more generally, for controlling a program in real-time. A class of meta-instruments allows to control the activation and release of notes. Interpretation of musical pieces based on activating and releasing notes has been studied by Haury [39].

Haury identifies four ways for interpretation: *dynamic variations* as the possibility to continuously modify the volume of the notes during the performance, *accentuation* as temporary volume variations, *phrasing* as modifying the binding of the notes, and *agogic variations* as the possibility to change the date of beginning and end of the notes. Haury's research focuses on agogic modifications. As examples of agogic modifications, in Haury's meta-instrument, the *metapiano*, the musicians can start or stop a group of notes through control points placed in the piece that he calls *interaction points*. A pause is a good example of interaction point in instrumental music because the musician or the conductor can choose the duration of the pause. Haury's work inspired Allombert *et al.*'s models of interactive scores.

## 5.4 Score following

Another kind of systems capable of real-time interaction are *score following* systems [23]. To use such systems, we must first write a score for the musician and for the computer. During execution, such systems track the performance of a real instrument and they may play music associated to certain notes of the piece. Nevertheless, to use these systems it is necessary to play a music instrument; whereas to use interactive scores, the user only has to control some parameters of the piece, such as the starting and ending times of the temporal objects. Score following systems can also provide temporal relations and hierarchical relations [30]; however, the system tracks the performance of a music instrument and is not meant to work with a meta-instrument. In contrast, one of the main advantages of interactive scores is meant to work with meta-instruments.

## 5.5 Asynchronous dataflow languages

Stream processing can be modeled as a collection of separate but communicating processes. Dataflow is the canonical example of stream processing. There is *synchronous dataflow* and *asynchronous dataflow* [88]. Synchronous dataflow they lack of FIFO queues to communicate channels like asynchronous dataflow languages. This is a main difference between the synchronous and asynchronous dataflow languages.

As an example, asynchronous dataflow languages *Max/MSP* and *Pure Data (Pd)* [70] are often used to control signal processing and control events by human gestures. Max and Pd distinguishes between two levels of time: the *event scheduler* level and the *digital signal processor (DSP)* level. Max and Pd programs, called *patches*, are made by arranging and connecting building-blocks of objects within a visual canvas. Objects pass messages from their outlets to the inlets of connected objects. The order of execution for messages traversing through the graph of objects is defined by the visual organization of the objects in the patch itself[21].

---

21 http://en.wikipedia.org/wiki/Max_(software)

There are several problems with Max and Pd that we aim to overcome. First, their schedulers control both audio signals and control messages together and they do not support parallelism, thus they often fail to deliver control messages at the required time; for instance, when they work under a high CPU load, which is common when they process video, 3D images and sound. We present some insights on how to solve this problem; nonetheless, this is still an open problem.

To solve the scheduling problem and to write high-performance DSPs for Max and Pd, users often write C++ plugins to model loops and independent threads. C++ plugins solve part of the problem, but the control messages –for the input and output of these plugins– are still being scheduled by Max or Pd's schedulers.

Second, there is another problem with Max and Pd: they do not provide an environment to design scenarios. The different time scales are often unrelated and cannot be controlled in the same tool: *Discrete user gestures* (e.g., clicking the mouse), *control events* (e.g., control messages) and *signal processing* have different sampling frequencies and computing models.

One goal of the extension of interactive scores with signal processing is to overcome the existing problems of the asynchronous dataflow languages mentioned.

## 5.6 Synchronous dataflow languages

There are three well-known french *synchronous languages*: *Esterel*, *Lustre* [37, 38] and *Signal* [33]. Benveniste *et al.* discussed the advantages and limitations of such languages 12 years after they were conceived [17]. They argue that synchronous languages were designed to implement real-time embedded applications, thus such languages work on the deterministic concurrency paradigm and they are meant to model deterministic system behavior. Synchrony divides time into discrete intervals and supposes that operations take no time (e.g., to assign a variable or read a value).

Benveniste *et al.* argue that Esterel is imperative and it is well-suited for describing control. Signal is based on the reactive programming paradigm: A program does something at each reaction and it may be embedded in some environment. Signal is a multiclock language. Lustre supports recursive definitions, but may not contain cyclic definitions, and a variable can only depend on past values. Both Lustre and Signal have clocks to align streams, but they lack of FIFO queues to communicate channels like asynchronous dataflow languages. This is a main difference between the synchronous and asynchronous dataflow languages.

A very useful feature of synchronous dataflow languages is multirate computation. Using multirate computation, it is possible to easily handle control signals, video signals and audio signals that have different sampling rates. In fact, Forget compared the mutirate capabilities of Esterel, Lustre and Signal [31]. Forget argues that in Lustre each variable is a flow. Lustre has a clock, but multirate is hard to describe. In Signal, variables are signals instead of flows. Clocks in Signal are first class objects; therefore, it can be polychronous, but multirate is also hard to achieve. Finally, Esterel focuses on control flow, where several modules communicate through signals, Esterel also has some asynchronous extensions and automated verification, but does not support multirate.

*Faust* is a synchronous language with formal semantics for multirate; however, this functionality has not yet been implemented [41]. Faust is a functional programming language for signal processing. In Faust, DSP algorithms are functions operating on signals. Faust programs are compiled into efficient C++ code that can be used in multiple programming languages and environments; for instance, in Pure data [35]. Faust is the DSP language we chose for our extension of interactive scores with signal processing.

There is another well-known synchronous dataflow language. *Csound*[22] has three types of variables with different time levels (and different sampling rates): instrument variables, control variables and audio variables. In fact, control variables correspond to event scheduler sampling rate and audio processes to DSP level in Max. Nonetheless, Csound does not provide sophisticated mechanisms to temporally relate instrument, control and audio variables; for instance, to say that one microsecond after an audio signal reaches a peak, a control variable changes its value, causing three instruments to play a note whose duration is the distance between such peak and the last peak the audio signal reached.

---

22 http://www.csounds.com/

## 5.7 Process calculi

Process calculi (or process algebras) are a diverse family of related approaches to formally model concurrent systems. Process calculi provide high-level description of interactions, communications, and synchronizations between a collection of independent processes. They also provide algebraic laws that allow process descriptions to be manipulated and analyzed, and permit formal reasoning about equivalences between processes; for instance, using *bisimulation* [82]. Intuitively, two systems are bisimilar if they match each other's moves. In this sense, each of the systems cannot be distinguished from the other by an observer. A well-known process calculus is the pi-calculus. Unfortunately, the pi-calculus is not well suited to model reactive systems with partial information.

*Concurrent constraint programming* (CCP) [83] is a process calculus to model systems with partial information. In CCP, a system is modeled as a collection of concurrent processes whose interaction behavior is based on the information (represented by constraints) contained in a *global store*. Formally, CCP is based on the idea of a *constraint system*. A constraint system is composed of a set of (basic) constraints and an entailment relation specifying constraints that can be deduced from others.

Although constraint systems suppose a big flexibility and modeling power for concurrent systems, Garavel argues that models based on process calculi have not found widespread use because there are many calculi and many variants for each calculus, making difficult to choose the most appropriate [32]. In addition, he argues that it is difficult to express an explicit notion of time and real-time requirements in process calculi. Finally, Garavel argues that existing tools for process calculi are not user-friendly and there are not many tools available.

A position in favor of process calculi is defended by Olarte *et al.* [60, 59]. They showed that CCP calculi have been used in several applications such as music interaction, security protocols and systemic biology. They explained that CCP has different variants to model mobility, probabilistic behavior, hybrid systems, discrete time and real-time.

We also argue, in favor of CCP, that there has been a growing interest for CCP models of music interaction in the last decade [77, 80, 78, 79, 75, 5, 100, 58, 59, 105]. CCP processes can be analyzed from both a behavioral and declarative point of view, making them suitable for simulation and for verification of properties. Some programming languages have also been developed following the concepts of CCP. As an example *Mozart/Oz* [74, 113] is a multiparadigm programming language inspired in the CCP paradigm.

Process calculi has been applied to the modeling of interactive music systems [104, 111, 99, 110, 10, 106, 100, 59, 97, 93, 95, 98, 11, 103, 94, 101, 102, 92] and ecological systems [107, 64, 109, 65, 108].

Although there are programming languages based on CCP, as Garavel argued, the explicit notion of time is missing in most process calculi and, unfortunately, it is also the case of CCP. In CCP it is not possible to delete nor change information accumulated in the store. For that reason, it is difficult to perceive a notion of discrete time, useful to model reactive systems communicating with an external environment (e.g., motion sensors and speakers).

The temporal concurrent constraint (*tcc*) [84] calculus circumvents this limitation by introducing the notion of discrete time as a sequence of *time units*. At each time unit, a CCP computation takes place, starting with an empty store (or one that has been given some information by the environment). In fact, tcc has been shown to be very expressive to model synchronous languages such as Lustre and Esterel [91]. There is also an interpreter to execute tcc models [89].

The non-deterministic timed concurrent constraint (*ntcc*) [56] adds non-determinism and asynchrony to tcc. Ntcc has been extendedly used for musical applications. We chose ntcc to express operational semantics of interactive scores because it allows for verification of temporal properties; for instance, it has been used to model music improvisation systems and a western-african music problem [77, 78]. In addition, there is a real-time capable interpreter for ntcc [100], and verifications tools and techniques are being developed in the recently started Colciencia's REACT+ project[23]. Finally, another advantage of ntcc is that it handles very naturally temporal constraints.

## 5.8 Temporal constraints

Temporal constraints have gained interest among scientists ever since the invention of artificial intelligence. Temporal constraints are often used for temporal planing of autonomous robots. Lately, the music interaction community developed an interest on temporal constraints for the design of interactive music.

There are two well-known types of temporal constraints: *metric (or quantitative) constraints* and *qualitative constraints*. Metric constraints restrict the distance between points and qualitative constraints are relative positions. A metric constraint is, for instance, "a point occurs five time units after another", and a qualitative constraint is, for instance, "a point occurs strictly before another".

There are some well-known classes of qualitative constraints: *interval-interval* (also known as *Allen's relations* [1], shown in Figure 13), *point-to-point* and *point-interval*. Interval-interval temporal relations were conceived to model dense (continuos) time, but they can also be used for discrete time. According to Gennari, point-to-point are more expressive than point-interval relations when interval-interval does not include disjunction. When interval-interval temporal relations include disjunctions, they are more expressive than the other classes, but its satisfiability is NP-Hard [34].

There are also some well-known classes of quantitative constraints: *unary constraints* and *binary constraints*. They express location and distance respectively, both concepts important in music, but useless without the concept of relative positions.

---

23 REACT+ is a colombian project supported by *Colciencias* to develop verification and simulation tools for ntcc calculi. http://cic.javerianacali.edu.co/wiki/doku.php?id=grupos:avispa:react-plus.

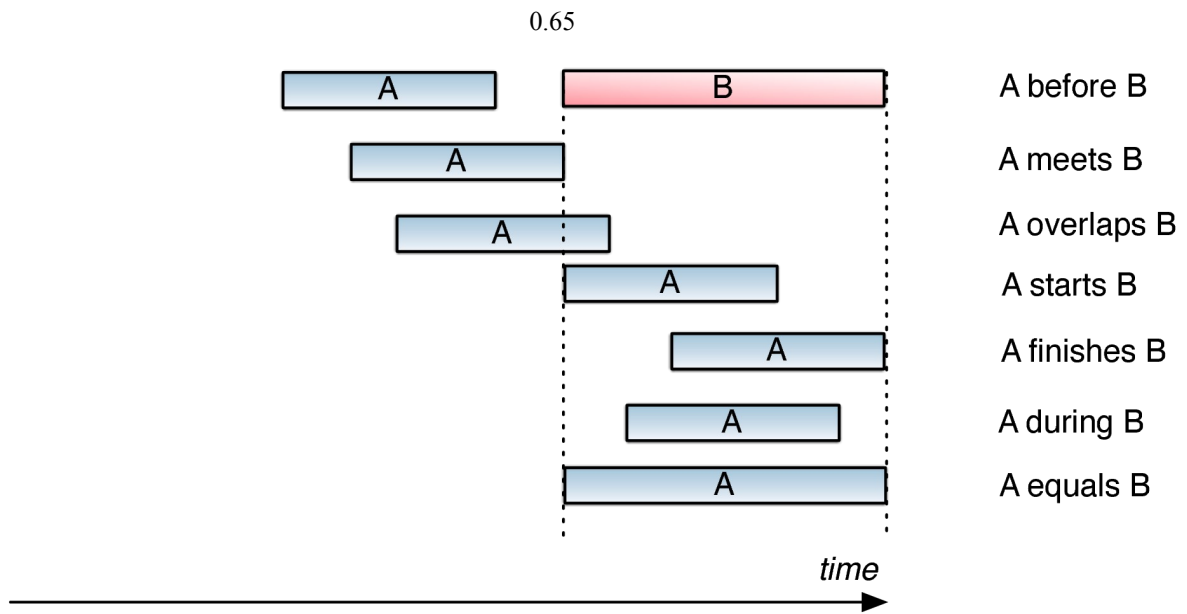

Figure 13: Allen's interval-to-interval relations.

Fortunately, Meiri formalized a new class: the combination of both qualitative and metric constraints [51]. Meiri's approach is simple: qualitative constraints can be represented as quantitative constraints; for instance, the relation < can be represented by the interval $(0, \infty)$. A subset of Meiri's new class can be represented as a *simple temporal problem* [27] when each temporal constraint is given by a single interval. In interactive scores, we combine point-to-point qualitative relations with unary and binary quantitative constraints, as proposed by Meiri.

There is another kind of temporal constraints: *hierarchical unification-based temporal pattern grammar* [20]. The unification-based temporal grammar is meant to describe multivariable time series. Such a grammar is an extension of context-free grammars with *Prolog clauses* evaluated as side conditions. Temporal patterns use logical disjunction and they have been successfully applied to the recognition of sleeping disorders. It has also been used to data mining hierarchical temporal patterns in multivariable time series. Nonetheless, Biundo *et al.*'s temporal grammar is not meant for real-time operations.

# 6 Conclusions

We described sequencers which are software to design music interaction. Sequencers are usually based on a fixed timeline or on cue lists. Some software provide both time models but they are temporally unrelated. An advantage of interactive scores is to relate temporally both time models and to model conditional branching.

There is also hardware to control music interaction. Meta-instruments are musician-machine gesture transducers intended for controlling a program in real-time. As an example, a meta-instrument can control the start and end of groups of notes, allowing for the interpretation of complex pieces with interfaces as simples as a one-touch piano. This work inspired the first models of interactive scores. In contrast, there are score-following systems, in which a real-instrument is needed. A score-following system tracks the performance and plays electronics associated to the notes of the score.

There are also synchronous and asynchronous dataflow paradigms, which are paradigms closely related to interactive scores. Asynchronous dataflow is meant to handle asynchronous events such as user interactions, whereas synchronous languages are meant to design real-time applications and they are based on a model of deterministic concurrency. Heterogeneous systems are systems that combines several paradigms, for instance, asynchronous and synchronous languages.

Heterogeneous systems combining asynchronous and synchronous circuits can be designed using schemes such as global asynchronous, locally synchronous. A special case, of interest for interactive scores is called *ratiochronous*, in which the receiver's clock frequency is an exact multiple to the sender's, and both are derived from the same source clock. This design scheme could be useful to synchronize interactive scores with a signal processing system, but also with other systems such as a score following system.

Process calculi are approaches to formally model concurrent systems. As an example, ntcc describes partial information by the means of constraints, it provides discrete time units, and it models asynchrony and non-determinism. Ntcc has been used in the past to model interactive scores. It handles naturally temporal constraints. A similar approach is Petri nets, which is is another model of concurrency with an intuitive graphical notation. An extension of Petri nets with time and hierarchy has been used to model interactive scores in the past and for synchronization of music streaming systems. Ntcc has also been used to model interactive scores

There are other existing models of interactive scores. First models were conceived to control the starting and ending times of the notes of a score. They also included different temporal relations; for instance, to model two temporal objects that overlaps, by the means of Allen's relations. Later extensions included a Petri nets operational semantics. Finally, there are extensions of interactive scores with conditional branching. Note that the Petri nets semantics of interactive scores were implemented in an efficient C++ library called Iscore, and it is currently being used by i-score.

Most scenarios and musical pieces with interactive controls have no formal semantics. Interactive scores is a formalism to describe interactive scenarios based on temporal constraints. In this dissertation, we introduced an event structures semantics of interactive scores, we formalized some properties, and we proved that the event structures semantics complies with the temporal constraints of the score. With the event structures semantics, we expressed several properties about the traces of execution that are difficult to express and prove using constraints.

We introduced the *dispatchable event structures (DES)*: event structures whose temporal object durations and temporal distances among objects are integer intervals. DES can be dispatched online by relying only on local propagation: This is achieved by transforming the constraint graph into an *all-pairs shortest-path graph*; however, that drastically increases the number of arcs. In the future, we propose to minimize the number of arcs of such networks, as proposed by Muscettola *et al.* [55].

Although event structures provide a theoretical background to specify properties and understand the system, there is no difference between interactive objects and static temporal objects in the event structures semantics: such a difference can only be expressed in the operational semantics. This means that the event structure semantics are not *fully abstract* with respect to the operational semantics: Operational equivalence does not always coincides with denotational equality. It is an open issue how to capture the behavior of interactive objects in the event structures semantics.

Operational semantics are based on the dispatchable normal form of the event structures of the score. A score is in normal form when it does not have zero-duration event delays. The computation of the normal form is similar to the algorithm to transform a score into a Petri net proposed by Allombert *et al.* [2]: In Petri nets semantics of interactive scores, points of temporal objects executed at the same time share the same *place* (i.e., state). Other algorithms for optimization problems include [62, 53, 72].

## Comparison with Allombert *et al.*'s model.

We believe that this dissertation extends significantly Allombert *et al.*'s model because it provides a concise operational semantics for interactive scores whose temporal object duration can be any interval of integers. Allombert *et al.* proposed temporal relations with flexible intervals with only {0}, [0,∞) and (0,∞) intervals [5, 4]. In fact, arbitrary integer intervals are not allowed in neither Virage nor i-score, only flexible-time intervals. To handle temporal relations with arbitrary intervals, Allombert proposed in [2] to either build a *Hierarchical colored time stream Petri net*, adding a big number of new places (states), or to use a constraint store that is unrelated to the Petri nets semantics, and the combined semantics of Petri nets interacting with a constraint store are not given.

There is another disadvantage of Allombert *et al.*'s models: Temporal relations are limited to *Allen's relations*. Allen's relations do not allow to represent quantitative relations between two objects easily; for instance, "object *a* occurs 3 time units after object *b*". Using Allen's relations, it is neither possible to say "the start of object *a* is before the end of object *b*". These kind of relations are easily modeled using point-to-point temporal relations. In fact, recently, i-score has moved forward to point-to-point temporal relations.

A conditional branching extension was presented in [2], but no temporal relations were allowed. We struggled to allow temporal relations and conditional branching in the same model. As an example, it is possible to model conditions and also preserve temporal properties over all the branches, for instance, that . In our first models of conditional branching, published in [101, 102], we allowed branches starting in the same point have different durations. We left aside such an approach because it makes many scores incoherent and unplayable.

An advantage of our extension of interactive scores with conditional branching with respect to previous models of interactive scores, Pure Data, Max and Petri Nets is representing declarative conditions by the means of constraints. Complex conditions, in particular those with an unknown number of parameters, are difficult to model in Max or Pd. To model generic conditions in Max or Pd, we would have to define each condition either in a new patch or in a predefined library. In Petri nets, we would have to define a net for each condition.

A disadvantage of our conditional branching model is that the number of event conflicts increases exponentially with respect to the hierarchy depth. Fortunately, the hierarchy depth is usually not so big, thus we argue that we do not need a formalism that supports hierarchical constructions, such as hierarchical Petri nets or statecharts.

Using timed event structures with conflicts, it is possible to model conditional branching: the possibility to choose among different continuations of the piece based on the preferences of the musician. In addition, Langerak describes in [45] how to encode recursive processes into event structures; in fact, loops in the interactive scores could be encoded with such a technique. Unfortunately, conditional branching drastically increases the complexity of the system; for instance, a score may contain dead-locks. An alternative for automated verification is *constraint programming*; for instance, to verify the playability of a score and calculate the potential time positions of the points of the score. Nonetheless, once again, we argue that, for some properties, the notion of *trace* is more appropriate.

Another advantage of our event structure semantics and our operational semantics is that they can express *trans-hierarchical relations*: temporal relations between objects with different parents. Trans-hierarchical relations are not possible to model with hierarchical time stream Petri nets used by Allombert *et al.* These relations are useful; for instance, to model temporal relations between videos and sounds that are contained in different temporal objects, allowing to define temporal relations among different media.

A key issue of this dissertation is that we executed interactive scores in a efficient manner. We want to encourage the use of process calculi to develop reactive systems. For that reason, this research focused on developing real-life applications with ntcc and showing that our interpreter *Ntccrt* is a user-friendly tool, providing a graphical interface to specify ntcc models and compiling them to efficient C++ programs capable of real-time interaction in Max and Pure Data (Pd). We argue that using ntcc to model, verify and execute reactive systems decreases the development time and guarantees correct process synchronization, in contrast to the graphical patch paradigm of Max and Pd.

## Disadvantages of our models.

A disadvantage of most ntcc tools is the syntax to write the input. Previous attempts to write ntcc processes directly as C++ classes, Lisp functions or visual objects has been proven to be insufficiently user-friendly. A compiler to parse ntcc into C++ classes is the "missing link" to allow non-programmers to use the real-time capable interpreter for ntcc (Ntccrt) and the ntcc time-bounded model checker (ntccMC), and could be the base for other CCP tools.

There are some other problems to execute interactive scores with Ntccrt. First, To compute the event structures semantics, its normal form and the dispatchable form by hand is very difficult. In the future, this should be done automatically. Second, ntcc recursive definition cannot be translated directly to Ntccrt because their encoding is based on nested non-deterministic choices hard to simulate. In the future, variables should be treated differently; for instance, using variables that can change value from a time unit to another one. Unfortunately, there are other problems that Ntccrt must overcome. Third, one may argue that although we can synchronize Ntccrt with an external clock (e.g., a metronome object) provided by Max or Pure Data, this does not solve the problem of simulating models when the clock step is shorter than the time necessary to compute a time-unit. To solve this problem, Sarria proposed to develop an interpreter for the *real time concurrent constraint (rtcc)* [85] calculus, which is an extension of ntcc capable of modeling time units with fixed duration. The reader may find a further discussion on executing time units with fixed durations in [100].

One may also argue that interactive scores had little applicability because they do not allow to describe signal processors. In this dissertation, we also extended the formalism of interactive scores with sound processing and micro controls for sound processors. We present an encoding of the scenario into a ntcc model –executed using the real-time capable interpreter *Ntccrt*– and a Faust program. Both programs interact during the performance of the scenario. We show how some interesting applications can be easily modeled in the formalism and how they can be executed in Pure Data. Using Faust and Ntccrt, we achieved an efficient and real-time capable performance of a scenario –even under high CPU-load. Nonetheless, our final goal is to integrate Ntccrt and Faust in a standalone program.

There is an interesting framework to evaluate the expressiveness of interactive multimedia formalisms: *Janin's dimensions*. There are several dimensions in multimedia interaction, according to Janin[24]: *Abstraction* that represents the hierarchy of temporal objects, *time* that represents the causality and can be thought as the *logical implication*, *parallelism* that represents that two (or more) objects can be executed simultaneously and can be though as an *logical and*, *alternative* that represents conditional branching and can be though as a *logical or*. Finally, there are dimensions for *value* that represents, for instance, the value of the pitch, volume or pan. Janin's dimensions are represented in Figure 14.

---

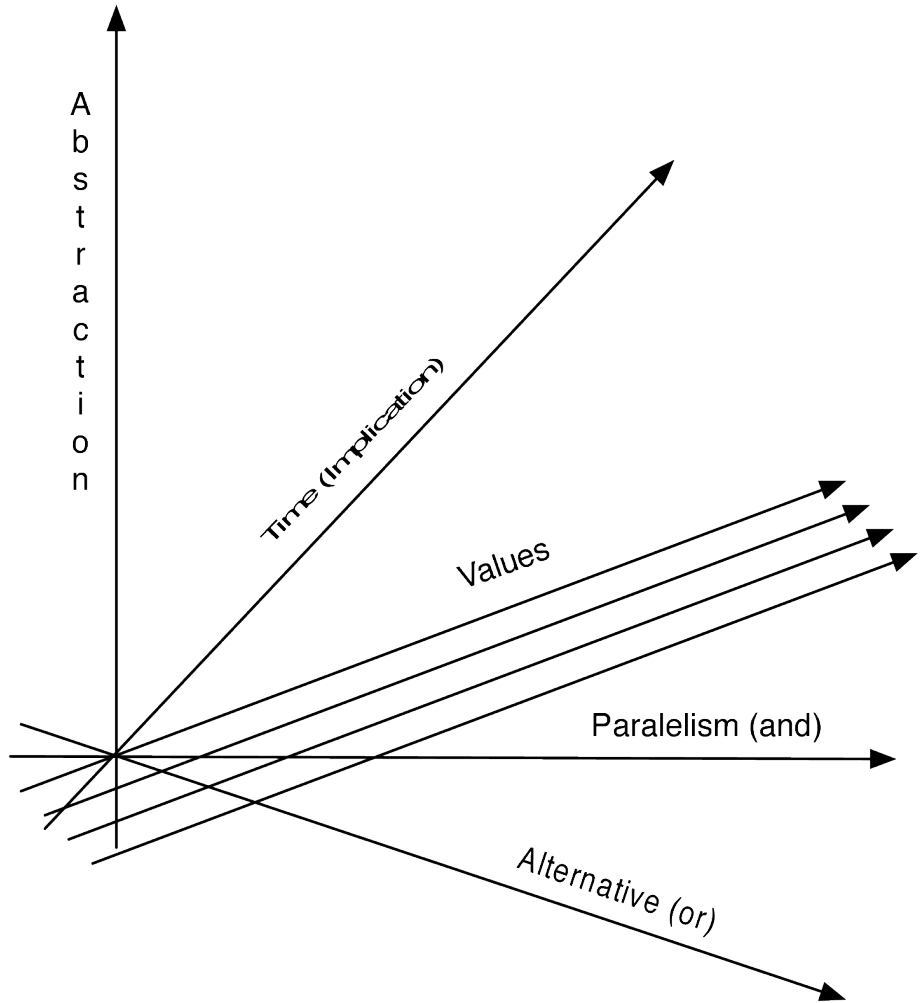

0.6

Figure 14: Janin's dimensions of interactive multimedia.

The hierarchical model of interactive scores allows us to express abstraction, time and values in the same two-dimensional space. In fact, i-score represents such interactive scores in a two-dimensional space. In the conditional branching model we can express abstraction, time, value and alternative, all in the same two-dimensional space because all branches starting on the same point have the same duration. Finally, in the signal processing extension, we can express time, value and parallelism in one two-dimensional space, and time, value and *dataflow* in another two-dimensional space. We argue that the dataflow dimension is missing among Janin's dimensions and should also be considered. The dataflow dimension describes how sound is transferred from one process to another. To represent time, value and dataflow together, we would need a tridimensional space; otherwise, arrows representing dataflow will overlap with those representing temporal relations.

## 6.1 Answers to problem statements

We have identified seven problems with existing software to design multimedia scenarios: (1) there is no formal model for multimedia interaction, (2) multimedia scenarios have limited reusability and difficulties with the persistence of multimedia scenarios, (3) time models (fixed timeline and cue lists) are temporally unrelated, (4) most multimedia interaction software products provide no hierarchy, (5) the different time scales are unrelated, (6) schedulers for multimedia scenarios are not appropriate for soft real-time, and (7) there is no model to combine temporal relations and conditional branching. In what follows we explain how the interactive scores formalism solves those problems.

First, interactive scores is a formalism to model multimedia scenarios. Event structures semantics allows to specify properties over the traces of execution. Ntcc semantics allows to understand the execution of the score and to specify temporal properties as well. Both semantics were proved to be related. Therefore, interactive scores is a formal model for multimedia interaction.

Second, scenarios described in interactive scores can be preserved because they have formal semantics. In addition, signal processors can be specified in Faust, which also has formal semantics. In fact, Faust can be used for preservation of music pieces because it provides formal semantics of all the audio processors used in the music piece [52, 15].

Third, time models are related temporally, for instance, we can specify that an object is executed strictly in the third second of execution, and we can also express that another object is executed between two and five seconds after the end of the previous object. Although, during the execution, micro controls are managed by Faust and macro controls by ntcc, it is also possible to express, for instance, that an object starts 500 microseconds after another, and it will end one second before another object.

Fourth, hierarchy is available in our model and it allows to constrain the execution times of the objects contained in another object.

Fifth, different time scales are available in our tool, but, unfortunately, they are temporally unrelated, as in many tools; for instance, it is not possible to relate the frequency of the clock that controls ntcc discrete time units to the signal processing sampling rate.

Sixth, the system is appropriate, even under high CPU-load, to interact with a human in real-time. The solution to this problem is relevant for the multimedia interaction domain because, in addition to sound processing, the computer may execute at the same time complex video and image operations. For that reason, we did the evaluation of our system under high CPU-load, obtained by executing several video processing operations concurrently.

Seventh, in interactive scores, it is now possible to combine conditions and intervals into a new type of relation called *time conditional relations*. In fact, by labeling these relations by `true` conditions, we can also express scores written in the pure temporal model. We managed to combine conditions and temporal relations by making the assumption that all branches starting in the same point have the same duration.

## 6.2 Future Directions

We propose some directions on the study and applications of interactive scores. Our final goal is to have a complete framework, as shown in Figure 15. The translation of conditional branching scores with loops into event structures is missing. In addition, operational semantics of conditional branching scores, for the general case, are missing. The translation of event structures semantics of scores with arbitrary durations into ntcc is also missing. Formal semantics of the integration of ntcc and Faust are missing. Some improvements for the model checker are missing to make it fully usable, and finally stand alone programs are missing to allow different applications of interactive scores, such as applications for music pedagogy. In what follows, we explain in detail some of these issues.

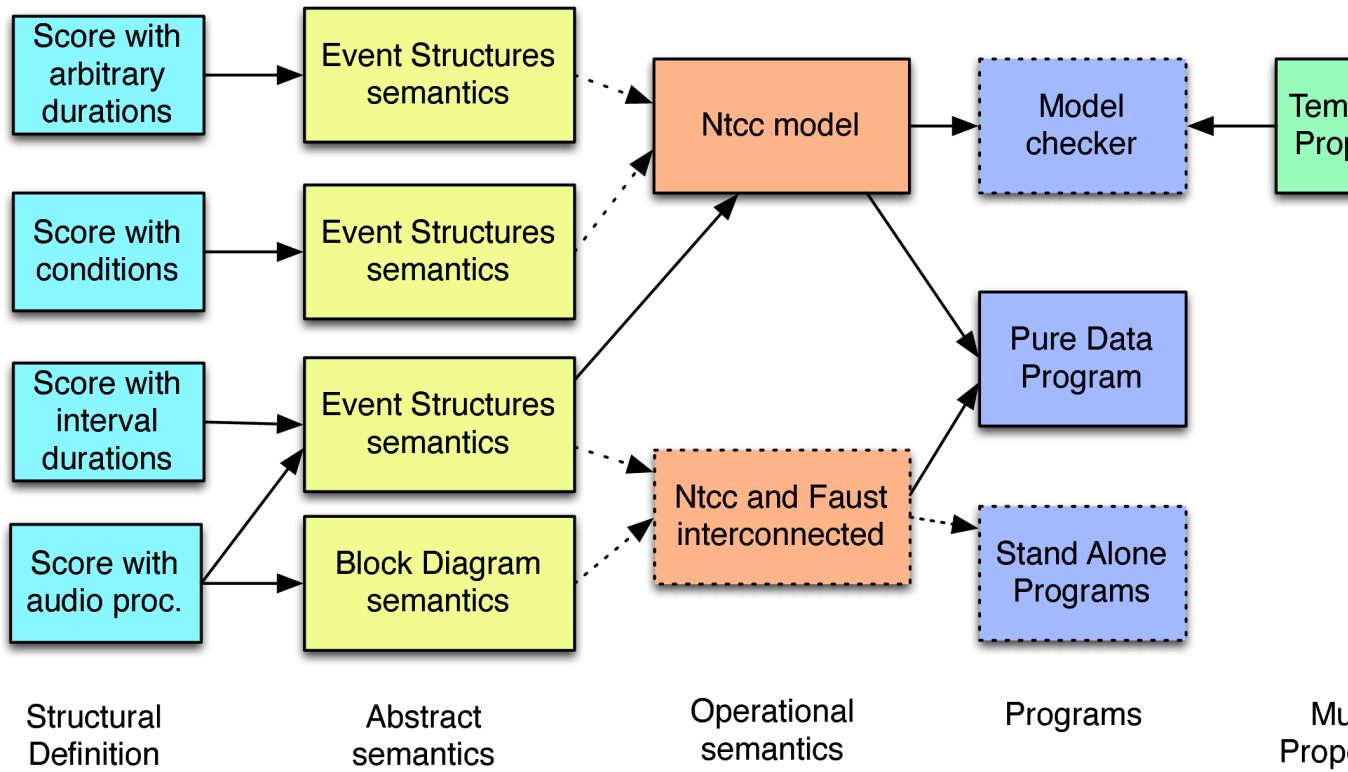

Figure 15: Diagram of the complete interactive scores framework. Dashed-arrows and dashed-lines represent translations, semantics and programs that are missing or are incomplete.

## Signal processing extension.

To improve the expressiveness of interactive scores, we should allow multiple points inside a temporal object, instead of just start and end points, as usual. Janin has already explained the advantages of such an approach to model rhythmical structures [40] .

We also propose to extend our implementation to handle audio files efficiently. *Libaudiostream*[25] is an audio library, developed at the french research institute *Grame*[26], to manipulate audio resources through the concept of streams using Faust programs. Including Libaudiostream in our framework, it will be possible to design a scenario where a temporal object loads a sound file into memory, filter it in Faust, and then, play the sound in Faust at the appropriate time. Precision is guaranteed because the time to load the file and to process it is foreknown in the scenario. Currently, we have to rely on third-party programs, such as Pd, to do handle audio files, and to communicate the control signals from Ntccrt to Faust.

It has been already discussed that Faust can be used to assure the persistence of music pieces with sound synthesis. We believe that such an approach could be used for the extension of interactive scores with signal processing. To solve that problem, Allombert developed a XML file format for interactive scores. This file format is currently used in Virage and i-score; however, it does not allow to represent the hierarchy, point-to-point temporal relations nor a set of possible durations of a temporal object.

In the future, we also want to to translate files from *music XML* and *music markup language* (*mml*) to our interactive scores XML format. We also want to represent scores with signal processors in our XML format.

## Conditional branching extension.

Event structures semantics for scores with loops is not easily defined because events can only be executed once; therefore, to define semantics we need infinite number of events, as proposed by Langerak in [45]. Afterwards, it will be required to translate such event structures semantics into operational semantics in ntcc with a finite number of processes.

## Automatic verification.

At the time of this writing, there are no formal semantics of a heterogeneous system that synchronizes concurrent objects, handles global constraints, and controls audio and video streams. Modeling this kind of systems will be useful in other domains such as *machine musical improvisation* and *music video games*. An advantage over the existing implementations of these systems will be verification.

We believe that any Faust program could be translated into ntcc based on the results obtained by Rueda *et al.* in [79]. Rueda *et al.* translated the Karplus-Strong Faust program into ntcc. Although it is clear that the execution of a Ntccrt simulation cannot be done at the sound processing sampling frequency, such a translation could be used to verify properties of correctness of a scenario where ntcc and Faust interact (e.g., playability).

In the proof system of ntcc, we could prove properties like "10 time units after the event is launched, during the next 4 time units, the stream $B$ is the result of applying a *gain filter* to the stream $A$"; however, real-time audio processing cannot be implemented in Ntccrt because it requires to simulate 44100 time units per second to process a 44.1 kHz sound. If we replace some ntcc processes by Faust plugins, we can execute such a system efficiently, but we cannot verify that the properties of the system hold. There is one open issue: How to prove that a Faust plugin that replaces a ntcc process obeys the temporal properties proved for the process. We discussed this issue in [96].

A first step to achieve the goal explained above is our model checker for ntcc, ntccMC. In ntccMC, we provide a prototype of a parser for ntcc syntax, but the parser can be improved. As an example, build an efficient representation of the process hierarchy, instead of a directed tree, so that two equivalent processes do not have to be encoded twice.

There is another disadvantage of *ntccMC*: Although FSA operations have lower complexity than operations over Büchi, the implementation needs to be improved to be used in bigger examples. The hash-table based automata class, provided by the *automata standard library*, is parametrized, during compilation time, by the size of the alphabet which is the number of relevant constraints. In addition, the number of relevant constraints is bounded by $n!$, where $n$ is the number of constraints that appear in the process and the formula. In addition to having a factorial number of constraints, constraint deduction is based on search, thus the domains of the variables should not be too big to be tractable.

## Scores whose temporal objects have arbitrary durations.

This extension will allow us to represent rhythmical patterns using temporal objects. When the duration of a temporal object can be an arbitrary set of integers, we can model rhythmical patterns; for instance, that a music object should be played at beats one, three or five (but not two nor four). Constraints of this form are found in the improvisation system presented by Rueda and Valencia in [78].

**Example 1** *As an example, Figure 16 is a score to represent rhythms. Object a's start time could be in the 1st,3rd,5th,9th or 12th time unit and its duration could be 1,2,3 or 4 time units.*

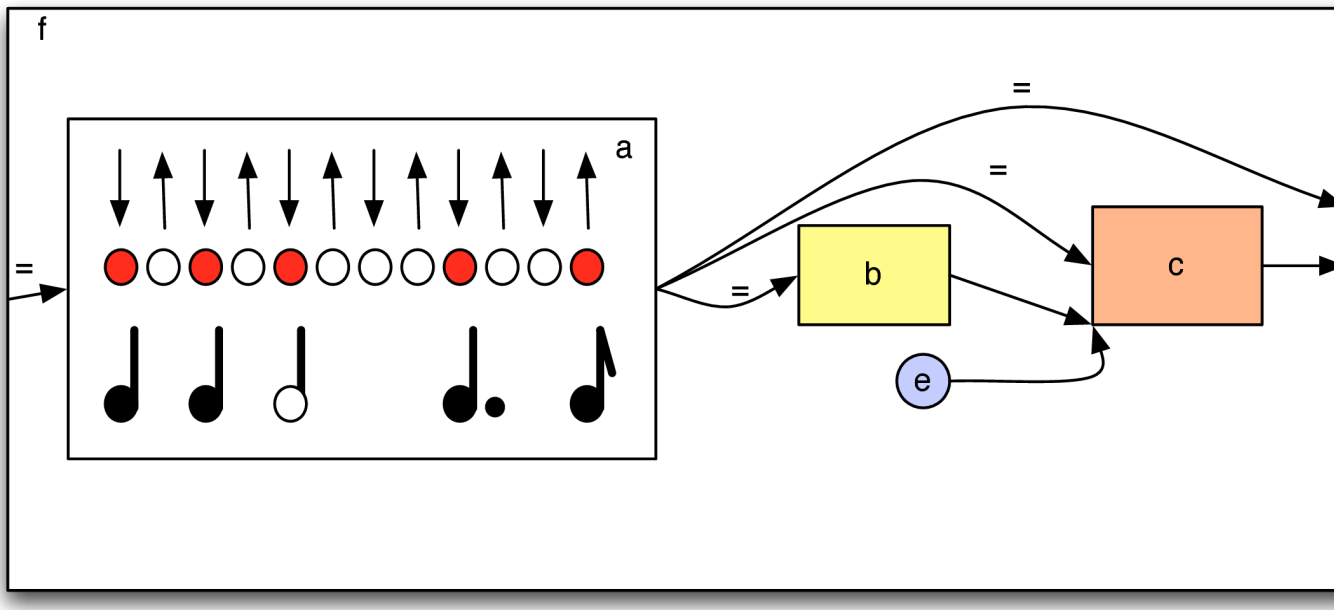

*Figure 16: A scores whose temporal objects have arbitrary durations.*

The satisfiability of a score with this kind of temporal constraints is equivalent to a disjunctive temporal problem, which is well-known to be NP-complete. One alternative to cope with this problem is to do a static analysis; for instance, a space efficient backtrack-free representation for constraint satisfaction problems [16]; however, to achieve such as representation, the order on which the temporal objects are going to be executed must be foreknown. Nonetheless, there are some scores in which this is possible, but for many other it is not possible.

Another possibility to cope with this problem in real-time could be an extension of Truchet's approach to solve music constraint satisfaction problems with local search [112]. Nonetheless, her algorithm requires random initialization of the variables and iterative refinements. Such a random initialization could be an incoherent representation of the temporal objects in the timeline; for instance, an end point could be executed before a start point.

## Pedagogic applications.

There are several possible pedagogic applications that can be developed using interactive scores. One alternative is to use interactive scores for rhythmic exercises for music students, easily modeled by constraints. Anders *et al.* have already discussed this approach [63], but we believe that it could be improved by allowing user interactions and temporal relations, which is possible in interactive scores.

Another possibility is using user gestures to generate Electroacoustic music for pedagogical purposes. This was not possible before in interactive scores due to the lack of a signal processing extension. In the future, we could imagine scenarios, as those proposed by Kurtag *et al.* [44].

Finally, another possibility for future work is to use automatic generated fingering for piano or guitar to generate scores in which only "easy" playable notes (according to a fingering analysis) are played by the user and the "hard" playable notes are played by the computer. Note that automatic generation of piano fingering has been already studied by Robine, who also describes several related work on that subject [73].